\documentclass[prd,twocolumn,showpacs,nofootinbib]{revtex4}
\usepackage{amsmath}
\usepackage[dvips]{graphicx}
\usepackage{subfigure}

\newcommand{\half}{\frac{1}{2}}
\newcommand{\braket}[1]{\mbox{$\langle #1 \rangle$}}

\DeclareMathOperator{\arctanh}{arctanh}

\begin{document}
\preprint{}
\title{Kinks in the Hartree approximation}

\author{Mischa~Sall\'e}
\email{mischa.salle@helsinki.fi}
\affiliation{Helsinki Institute of Physics, P.O.Box 64,\\
FIN-00014 University of Helsinki, Finland}    

\date{\today}

\begin{abstract}
The topological defects of $\lambda \phi^4$ theory, the kink and antikink, are
studied in the Hartree approximation. This allows us to discuss quantum effects
on defects in both stationary and dynamical systems. The kink mass is
calculated for a number of parameters, and compared to classical, one loop and
Monte Carlo results known from the literature. We discuss the thermalization of
the system after a kink-antikink collision. A classical result, the existence of
a critical speed, is rederived and shown for the first time in the quantum
theory. We also use kink-antikink collisions as a very simple toy model for
heavy ion collisions and discuss the differences and similarities, for example
in the pressure. Finally, using the Hartree Ensemble Approximation allows us to
study kink-antikink nucleation starting from a thermal (Bose-Einstein)
distribution.
On a qualitative level, our results show only few dissimilarities with the
classical results, but on a quantitative level there are some important
differences.
\end{abstract}

\maketitle

\section{Introduction}

In recent years the increasing numerical power of computers has made it possible
to follow quantum fields in real time, studying phenomena such as
thermalization, phase transitions and the formation of topological defects.
Until recently this was mainly done using the classical approximation, which can
be justified when occupation numbers are large. However, this is never the case
for all momentum modes, and for systems far from equilibrium, this approximation
may also break down. The Hartree and large-$n$ approximations do capture quantum
effects but, until recently, the available numerical power did not allow for the
simulation of inhomogeneous systems, thereby excluding the study of many
physically interesting phenomena such as topological defects. It also has problems
with describing thermalization, due to the lack of direct scattering. Recently,
approximations based on an expansion of the $2PI$ effective action (see
Ref.~\cite{AaAh02} and references therein) and also on a truncation of
Schwinger-Dyson equations (see Ref.~\cite{CoDa02} and references therein), which do
include direct scattering have been shown to lead to a quantum thermal
equilibrium. Unfortunately, these methods are still numerically expensive and
are not yet able to describe inhomogeneous systems. In two
papers~\cite{SaSm00a,SaSm00b} we introduced the Hartree ensemble approximation,
which tries to combine the good features of both the classical approximation,
which is able to thermalize (classically), and the Hartree approximation, being
a quantum approximation and thus free from the Rayleigh-Jeans divergencies. In
the Hartree ensemble approximation, the initial density matrix is expanded on a
(overcomplete) basis of Gaussian density matrices. Expectation values can then
be written as a weighted average of Gaussian expectation values. The equations
of motion are derived using the Hartree approximation in each of the
``realizations.'' The partial ``mean fields'' of these different
``realizations,'' i.e., the expectation values of the field in the different
Gaussian initial states, are typically inhomogeneous and the interaction of the
quantum modes with the Fourier modes of these ``mean fields'' leads to
approximately thermal distributions. However, since these ``mean fields'' are
typically inhomogeneous, topological defects can and will be present if the
symmetries of the theory allow for them. These will be the topic of this paper,
we will look at the $\phi^4$ theory in 1+1 dimensions and its defects, kinks,
and make a comparison between the classical theory and the Hartree
approximation.

The rest of the paper in organised as follows. We will, in the remainder of this
introduction, briefly review the equations of motion and our definition of
instantaneous particle number, used in studying the distribution function. In
Sec.~\ref{sec:clas_kink} we will derive the classical kink solutions, as this
will facilitate comparison with the Hartree approximation. Then, in
Sec.~\ref{sec:kink_rest}, we will discuss Hartree kink-antikink pairs, initially
at rest and compare with the classical approximation. A study of the Hartree
kink mass will be made. In Sec.~\ref{sec:coll_kink} kink-antikink collisions
will be studied. We will look at the critical speed: below a certain speed the
pair annihilates, while above they bounce back. This has been extensively
discussed for the classical theory, but not yet for the quantum theory. We will
study whether the system thermalizes after the collision and use the collisions
as a toy model for heavy ion collisions, looking at the central region and its
dependency on coupling and initial speed. As a final topic we will, in
Sec.~\ref{sec:kink_nucl}, look at thermal kink nucleation, starting from a
Bose-Einstein thermal ensemble, as in Ref.~\cite{SaSm00b}. The conclusions are
in Sec.~\ref{sec:conclus}.

\subsection{Equations of motion}

In this section we very briefly discuss the Hartree equations of motion for the
$\phi^4$ theory, as used in the separate ``realizations'' of the initial
ensemble. For more information see Refs.~\cite{SaSm00a,SaSm00b,SaSm02}.

The Lagrangian density for our 1+1 dimensional system is given by\footnote{We
use the metric $(-1,1)$.}
\begin{equation}
\mathcal{L} = - \frac{1}{2} \partial_\mu \phi \partial^\mu \phi - \frac{1}{2}
\mu^2 \phi^2 - \frac{1}{4} \lambda \phi^4,
\end{equation}
giving rise to the Heisenberg (operator) equations of motion
\begin{equation}
-\partial_\mu \partial^\mu \hat\phi + \mu^2 \hat\phi + \lambda \hat\phi^3 = 0.
\end{equation}
By introducing expectation values of products of field operators one obtains a
hierarchy of equations: the equation for the $n$-point function will contain
also the $(n+2)$-function. Unless the initial density matrix, which is used in
the calculation of the expectation values, is Gaussian, this hierarchy will be
infinite. The Hartree approximation truncates it self-consistently by imposing
Gaussianity, thus factorizing all $n$-point functions with $n>2$ into $1$- and
$2$-point functions. It is convenient to expand the $2$-point function in a set
of mode functions, which in the free theory are simply plane waves. This
expansion is most easily implemented in the field operator
itself\footnote{We assume a large finite lattice size $L$ and use finite size
notation, i.e., a discrete spectrum.}
\begin{equation}
\hat \phi(x,t) = \phi(x,t) + \sum_\alpha f_\alpha(x,t) \hat b_\alpha + h.c.
\end{equation}
It is important to note that, due to the Hartree approximation, the creation and
annihilation operators $\hat b_\alpha$ and $\hat b^\dagger_\alpha$ can be chosen
time-\emph{in}dependent. They also satisfy the usual commutation relations. The
complex mode functions $f_\alpha(x,t)$ form a complete and orthogonal basis with
a Klein-Gordon-type inner product.
However, the $\alpha$-label in general does not coincide with the Fourier label
in time-dependent systems.

The Heisenberg equations of motion in the Hartree approximation can thus be
written as
\newlength{\boxsize}
\settowidth{\boxsize}{$f_\alpha$}
\begin{subequations}
\begin{align}
   \ddot{\phi} & = \Delta \makebox[\boxsize]{$\phi$} -
    [\mu^2 + \hphantom{3}\lambda \phi^2 +
    3 \lambda C ] \phi, \label{eq:phieom} \\
   \ddot{f}_{\alpha} & = \Delta \makebox[\boxsize]{$f_{\alpha}$} -
   [\mu^2 + 3\lambda \phi^2
   + 3 \lambda C ] f_{\alpha}, \label{eq:modeeom}
\end{align}
with
\begin{equation}
C = \sum_{\alpha} (2n_\alpha^0 + 1)|f_{\alpha}|^2, \qquad
n_\alpha^0 = \langle \hat{b}^\dagger_{\alpha} \hat{b}_{\alpha} \rangle.
\end{equation}
\label{eq:eom}
\end{subequations}
In $1+1$-dimensions, only the mass needs renormalization, due to the divergence
of the mode sum $C(x,t)$. We will do this by absorbing its divergent vacuum
part in $\mu^2$. In terms of the renormalized $\mu_\text{ren}^2$ we have
obtained a finite theory. In the rest of the paper we will only consider states
for which $n_\alpha^0=0$.

\subsection{Observables: Particle number}

In order to study thermalization we use the same definition for particle number
as before \cite{SaSm00a,SaSm00b,SaSm02}, using the $2$-point functions.
\begin{subequations}
\begin{align}
S(t,x-y) &= \overline{\braket{\hat\phi(t,x)\hat\phi(t,y)}} -
\overline{\braket{\hat \phi(t,x)}} \; \overline{\braket{\hat\phi(t,y)}},\\
U(t,x-y) &= \overline{\braket{\hat\pi(t,x)\hat\pi(t,y)}}
-\overline{\braket{\hat\pi(t,x)}} \; \overline{\braket{\hat\pi(t,y)}},
\end{align}
\label{eq:2pnt}
\end{subequations}
where the overline denotes an average over the center of mass coordinate and
optionally a time interval and/or ensemble of initial conditions:
\begin{multline}
\overline{ \braket{ \hat\phi(t,y+z) \hat\phi(t,y) } } = \\
\frac{1}{N} \sum_\text{ens.} \frac{1}{L\delta}\int_{t-\delta/2}^{t+\delta/2}
dt' \int_0^L dy \braket{ \hat\phi(t',y+z) \hat\phi(t',y) }, 
\end{multline}
where $N$ is the number of ensemble members, $L$ the lattice size and $\delta$ a
small time interval.
In principle the (asymmetric part of the) $\phi \pi$ correlation function could
also be used, but as it vanishes in equilibrium, we will refrain from doing
that.

We can split the $2$-point functions in the separate contributions from the mean
field and the mode functions: 
\begin{equation}
S(t,x-y) = S^c(t,x-y) + S^q(t,x-y), \label{eq:cq1}
\end{equation}
where
\begin{subequations}
\begin{align}
S^c(t,x-y) &= \overline{\phi(t,x)\phi(t,y)} 
- \overline{\phi(t,x)}\; \overline{\phi(t,y)},\\
S^q(t,x-y) &= \overline{C(t,x;t,y)}, \label{eq:cq2}
\end{align}
\end{subequations}
and similarly for $U(t,x-y)$. These expressions only depend on the relative
coordinate $x-y$. We take the Fourier transform
\begin{equation}
S_k(t) = \frac{1}{L} \int_0^L dz \; e^{-ikz} \, S(t,z),
\label{eq:Sk}
\end{equation}
with $k=(0, \pm 1, \pm2, \cdots)(2\pi/L)$.
Analogously to their expressions for a free theory we can use them to define
instantaneous particle numbers:
\begin{subequations}
\begin{align}
S_k(t) &= \left[n_k(t) + \half\right]\frac{1}{\omega_k(t)},
\label{eq:nomdef1} \\
U_k(t) &= \left[n_k(t) + \half\right]\,\omega_k(t).
\label{eq:nomdef2}
\end{align}
\label{eq:studef}
\end{subequations}
From these relations we can extract $n_k(t)$ and $\omega_k(t)$ as a function of
time. Compared to other definitions of particle number, this one has the
advantage of being manifestly real, thus making them particularly suitable for
studying out-of-equilibrium systems.

\section{Classical kink solutions\label{sec:clas_kink}}

When looking for stable solutions to the classical equations of motion of the
$\phi^4$ theory in an infinite volume, the field has to be in the vacuum at
spatial infinity in order to keep the energy finite. In the ``broken phase''
this still leaves some important freedom. For the trivial solution both ends
will be in the same vacuum. However, there is also a nontrivial solution, where
the field is in the different vacua in the different spatial infinities and hops
over the central barrier somewhere in between. This is the so-called (anti)kink
solution. The exact location of the kink does not influence the energy and an
infinite set of degenerate solutions exists, but transforming the kink
solution into the trivial solution would cost an infinite amount of energy, and
hence the topology has rendered it stable. See Ref.~\cite{Ra82} for a
standard introduction into the topic.

It is possible to calculate this lowest non-trivial solution using a very
elegant and general method, the Bogomol'nyi equation~\cite{Bo76}. In the broken
phase the classical static Hamiltonian (energy) is given by
\begin{align}
E &= \int dx \, \half (\phi')^2 + \frac{\lambda}{4} \bigl(\phi^2 - v^2\bigr)^2
\nonumber \\
&= \int dx \, \biggl\{ \half \biggl(\phi' \pm 
\sqrt{\frac{\lambda}{2}} \bigl(\phi^2 - v^2\bigr)\biggr)^2 \mp
\phi' \sqrt{\frac{\lambda}{2}} \bigl(\phi^2 - v^2\bigr) \biggr\} \nonumber \\
&= \int dx \, \half \biggl(\phi' \pm
\sqrt{\frac{\lambda}{2}} \bigl(\phi^2 - v^2\bigr)\biggr)^2 \nonumber\\
&\hphantom{=} \mp
\biggl[ \sqrt{\frac{\lambda}{2}} \bigl(\frac{1}{3}\phi^3 - v^2\phi\bigr)
\biggr|_{\phi(-\infty)}^{\phi(\infty)},
\label{eq:hamil_kink}
\end{align}
where $\phi'=\partial_x \phi$. The boundary term is equal to a finite constant
and determines the topological sector, trivial or nontrivial. For given
boundaries the energy is therefore bounded from below by the first term which is
larger than, or equal to zero, the so-called Bogomol'nyi
bound. The lowest energy state can be found by solving
the simple first order differential equation
\begin{equation}
\phi' = \mp \sqrt{\frac{\lambda}{2}} \bigl(\phi^2 - v^2\bigr),
\label{eq:bogomol}
\end{equation}
which has solutions
\begin{equation}
\phi = \begin{cases}
\pm v \, \tanh\Bigl[\frac{m}{2} (x-x_0) \Bigr], \\
\pm v.
\end{cases} \label{eq:broken_sol}
\end{equation}
Here $x_0$ is the point where $\phi$ goes through $0$ and $m=\sqrt{2 \lambda
v^2}$ is the quasiparticle or ``meson'' mass in the broken vacuum. In the top
line, the $+$ solution corresponds to a kink, the $-$ solution to an antikink.
The width of the kink is inversely proportional to the mass.

\begin{figure}[tbp]
\includegraphics[width=0.232\textwidth]{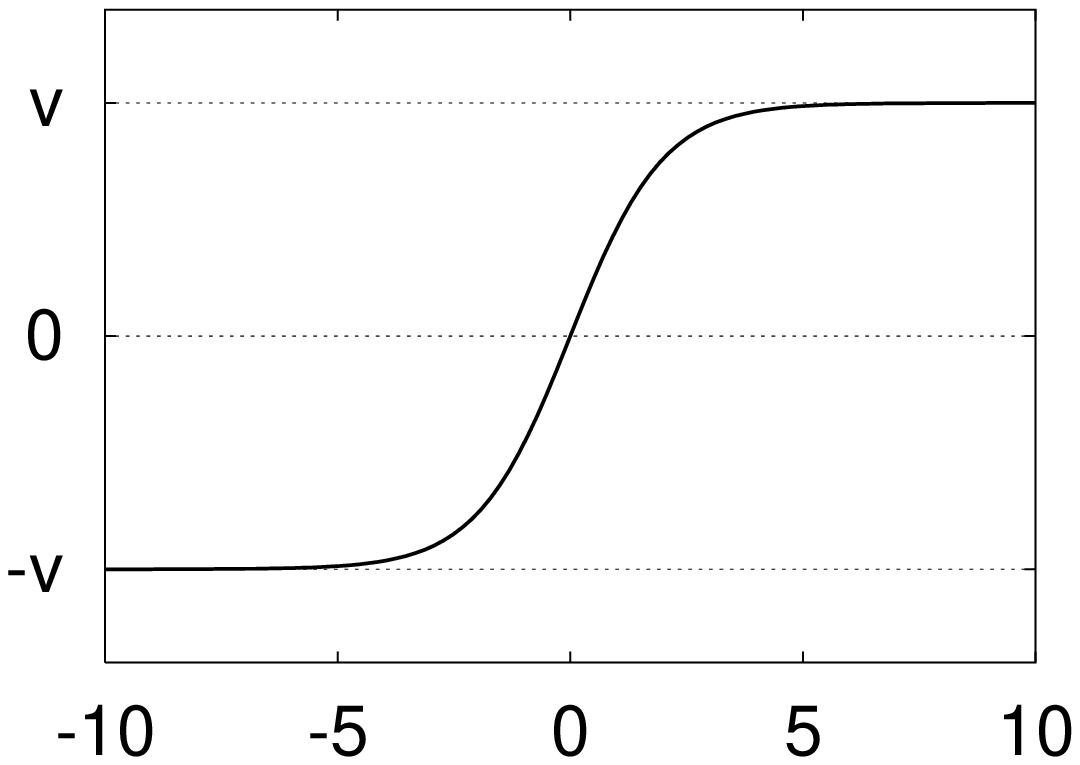}
\includegraphics[width=0.232\textwidth]{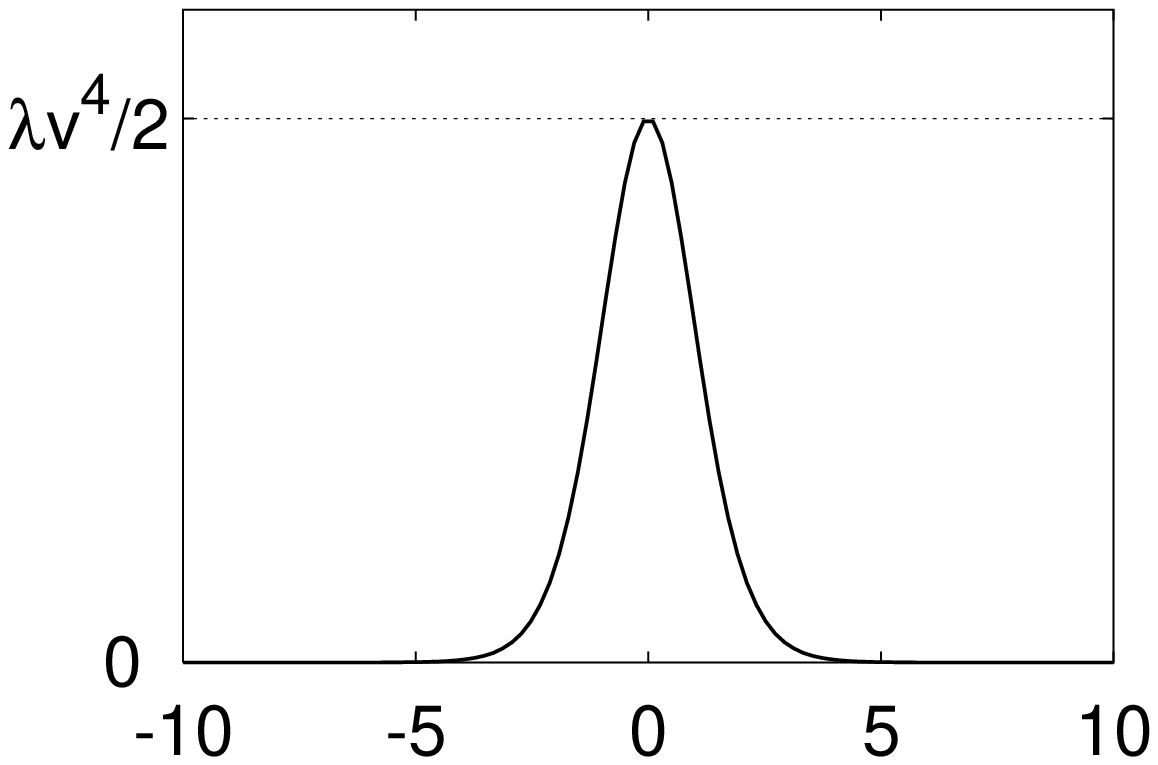}
\caption{Classical kink solution \eqref{eq:broken_sol} and its energy
density \eqref{eq:kink_edense} as a function of $x$ in units
$m^{-1}$.\label{fig:class_kink}}
\end{figure}
The kink energy density can be found from the top line of \eqref{eq:hamil_kink}
using \eqref{eq:bogomol}
\begin{equation}
e_\text{kink} = \frac{\lambda}{2} (\phi^2 - v^2)^2 =
\frac{1}{2} \lambda v^4 \Bigl(\tanh^2 \Bigl[\frac{m}{2}(x-x_0)\Bigr] -
1\Bigr)^2.
\label{eq:kink_edense}
\end{equation}
Both the solution \eqref{eq:broken_sol} and the energy density
\eqref{eq:kink_edense} are plotted in Fig.~\ref{fig:class_kink}.

The kink mass, defined as the energy of the static
kink or antikink, can be found either by integrating \eqref{eq:kink_edense} or
directly from the boundary term in
\eqref{eq:hamil_kink}
\begin{equation}
M_\text{kink} = \frac{2}{3} \sqrt{2 \lambda} v^3 = \frac{2}{3} m v^2 =
\frac{m^3}{3 \lambda}.
\label{eq:classkinkmass}
\end{equation}
Note that it diverges in the limit $\lambda \to 0$, showing the nonperturbative
nature of the configuration.

From the static kink solution \eqref{eq:broken_sol} we can easily find a moving
kink solution by a Lorentz
transformation:
\begin{equation}
\phi(x,t) = v \, \tanh\Bigl[\gamma \frac{m}{2} (x - x_0 - u t) \Bigr],
\end{equation}
where $\gamma=1/\sqrt{1-u^2}$. Note that this is not a solution of the
Bogomol'nyi equation \eqref{eq:bogomol}, since it is not static, however it is a
solution to the manifestly Lorentz covariant field equations. The total energy
of this moving solution is equal to $E_\text{kink} = \gamma M_\text{kink}$, as
expected for a relativistic particle. In summary, a Lorentz boost has the effect
of increasing the kink mass while decreasing its width.

\subsection{Lattice (arte)facts}

We end this section by mentioning some consequences of the use of a finite
lattice. 

The topological sector is determined by the type of boundary conditions: when
imposing periodic boundary conditions the net kink number is $0$, but a
kink-antikink pair can be considered. Although such a configuration is
inherently unstable, its survival time depends heavily on the relative distance,
as a result of the exponential attraction between them. When using antiperiodic
boundary conditions, the kink number is $\pm 1$. By shifting the kink over 1
lattice size $L$ through the boundary, the kink number changes sign.
This means one cannot distinguish between kink and antikink. However, the defect
itself is stable, as expressed by the nonzero kink number.

An unwanted side effect of the use of a lattice is radiation,
resulting from discretization errors and interfering with the kinks
themselves. Some solutions have been proposed, see for example Gleiser and
Sornborger \cite{GlSo00} and Speight and Ward \cite{SpWa94,Sp97,Sp99}. In
Ref.~\cite{GlSo00} a damping method is used to remove the radiation before it
can interfere. Speight and Ward \cite{SpWa94,Sp97,Sp99} propose a lattice
discretization preserving the Bogomol'nyi bound, thereby also removing the
radiation resulting from the discretization.
However, according to Ref.~\cite{AdAl01} this latter solution is only suitable
for free kinks, while in dynamical systems a standard discretization would be
simpler
with effectively the same accuracy.
We will not bother too much about this issue. If it will be necessary to
accurately follow an approaching pair
over a long period of
time, large lattices will be
used and the kink and antikink will be put close together initially: the
radiation is mainly emitted backwards and therefore reenters via the periodic
boundary condition. Thus, by setting
the soliton pair
close together, we can let it take a
long time for the radiation to reach
them again.

\begin{figure}
\includegraphics[width=0.483\textwidth]{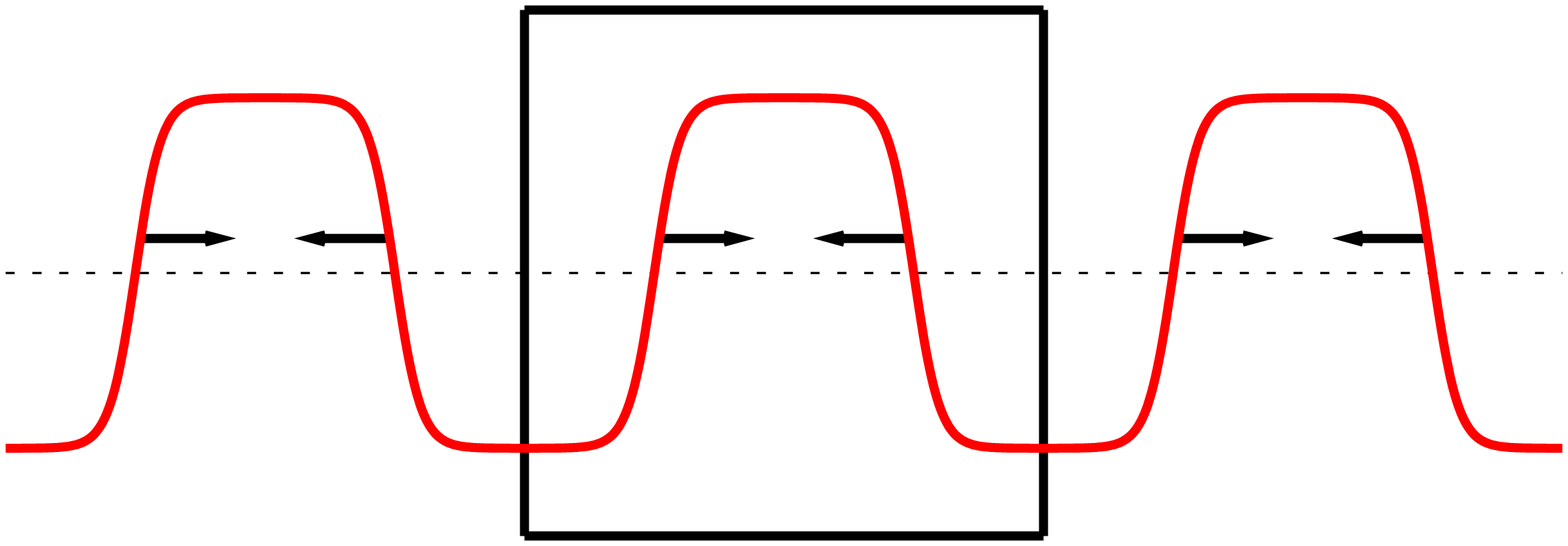}
\caption{Kink-antikink initialisation procedure.\label{fig:periodickink}}
\end{figure}
Finally, the use of (anti)periodic boundary conditions in a finite volume
introduces a small mismatch in the derivative of the field at the boundaries,
especially if the kink and antikink are close to them. We therefore use the
initialization setup sketched in Fig.~\ref{fig:periodickink}: we take several
configurations in several adjacent lattices and use the resulting configuration
in one of them, 
\begin{multline}
\phi_{K\overline{K}}(x,t)\Bigl|_{t=0,a_0} = -v + \\
\begin{aligned}
v \sum_{n=-N}^{N} \biggl\{
& \tanh \Bigl[ \gamma \frac{m}{2}(x+[x_0-ut]+n\cdot L)\Bigr] \\ - & 
\tanh \Bigl[ \gamma \frac{m}{2}(x-[x_0-ut]+n\cdot L)\Bigr] \biggr\},
\end{aligned}
\label{eq:iniclass}
\end{multline}
where $N$ is typically around $8$.

\section{Hartree kinks at rest\label{sec:kink_rest}}

Although interesting in itself the classical theory is not the full story and it
is therefore important to study quantized kink solutions. In
order to study dynamical processes,
such as
collisions, one has to use
approximation schemes, which are nonperturbative, real-time and which can
handle inhomogeneous configurations. For this purpose the inhomogeneous Hartree
approximation as studied in Ref.~\cite{SaSm00a} and \cite{SaSm00b}
should be particularly appropriate, being a nonperturbative semiclassical
approximation scheme.

\subsection{Initial condition}

Unlike in the classical theory, we cannot use the Bogomol'nyi argument to derive
a static soliton solution to the Hartree equations~\eqref{eq:eom}. However, if
the coupling is not too large, or more precisely, if the two vacua are well
separated and the barrier between them is high, the quantum corrections will be
relatively small. Furthermore, away from the physical position of the kink, the
field resides in one of the vacua, where an exact solution of the Hartree
equations is known. Therefore using the classical kink solution as initial
condition for the mean field, while using the free field plane wave solutions
for the mode functions will be close to a Hartree kink
solution, provided the coupling is not too close to the phase transition, i.e.,
the vacuum expectation value $v$ is (substantially) larger than $1$.
After the configuration has evolved, using the Hartree equations of
motion, from such an initial condition, it will, at least for a while, oscillate
around the actual stationary
solution.

A way to improve on this initial condition is to add a damping
term
\begin{equation}
- \Gamma \partial_t \phi
\label{eq:dampterm}
\end{equation}
to the right hand side of the mean field equation\footnote{Note that it is not
possible to do this at the level of the action or
Hamiltonian.}~\eqref{eq:phieom} and evolve from the aforementioned initial
condition to obtain an approximately stationary solution, which can then be used
as a new initial condition. We will discuss this setup further in
Sec.~\ref{sec:coll_kink}, when considering colliding kinks. In
Sec.~\ref{ssec:kinkmass} we will use a damping term to determine the quantum
kink mass.

Another possibility would be to start with the initial conditions as proposed in
Ref.~\cite{BoCo98}. However, these still are not stationary under the equations
of motion and the advantage is therefore only minor, so we choose to use the
simpler vacuum mode functions.

The numerical results presented in the next sections are all obtained using
the simple initial condition, with mode functions of free field form and
Eq.~\eqref{eq:iniclass} for the mean field.

\subsection{Numerical results: Static kink decay\label{ssec:kinkdecay}}

\begin{figure*}[tbp]
\begin{center}
\subfigure[Classical, $\lambda/m^2=1/1.25$. For clarity the $x$-coordinate is
shifted over $L/2$ through the periodic boundary
conditions.\label{fig:statclassstrong}]{
    \includegraphics[width=0.475\textwidth]{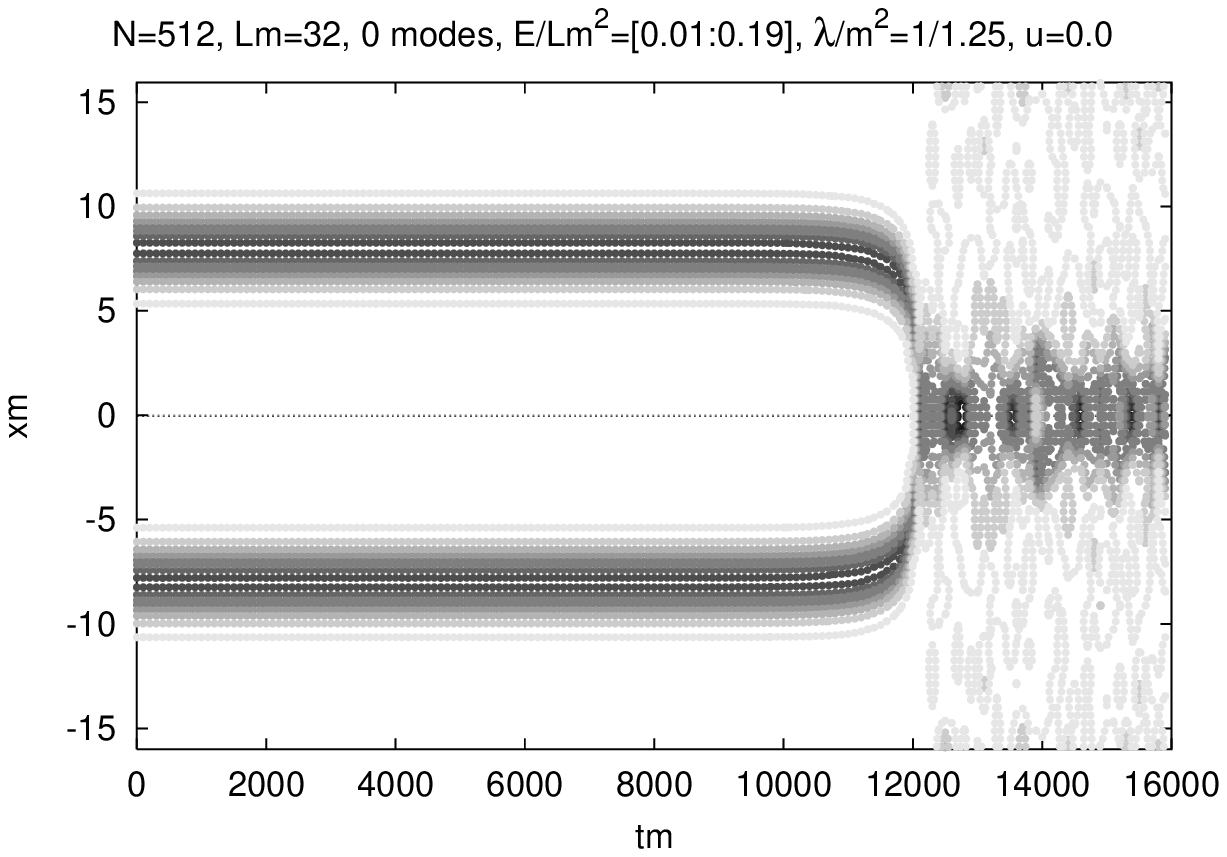}}
\subfigure[Hartree, $\lambda/m^2=1/1.25$.\label{fig:statHartstrong}]{
    \includegraphics[width=0.475\textwidth]{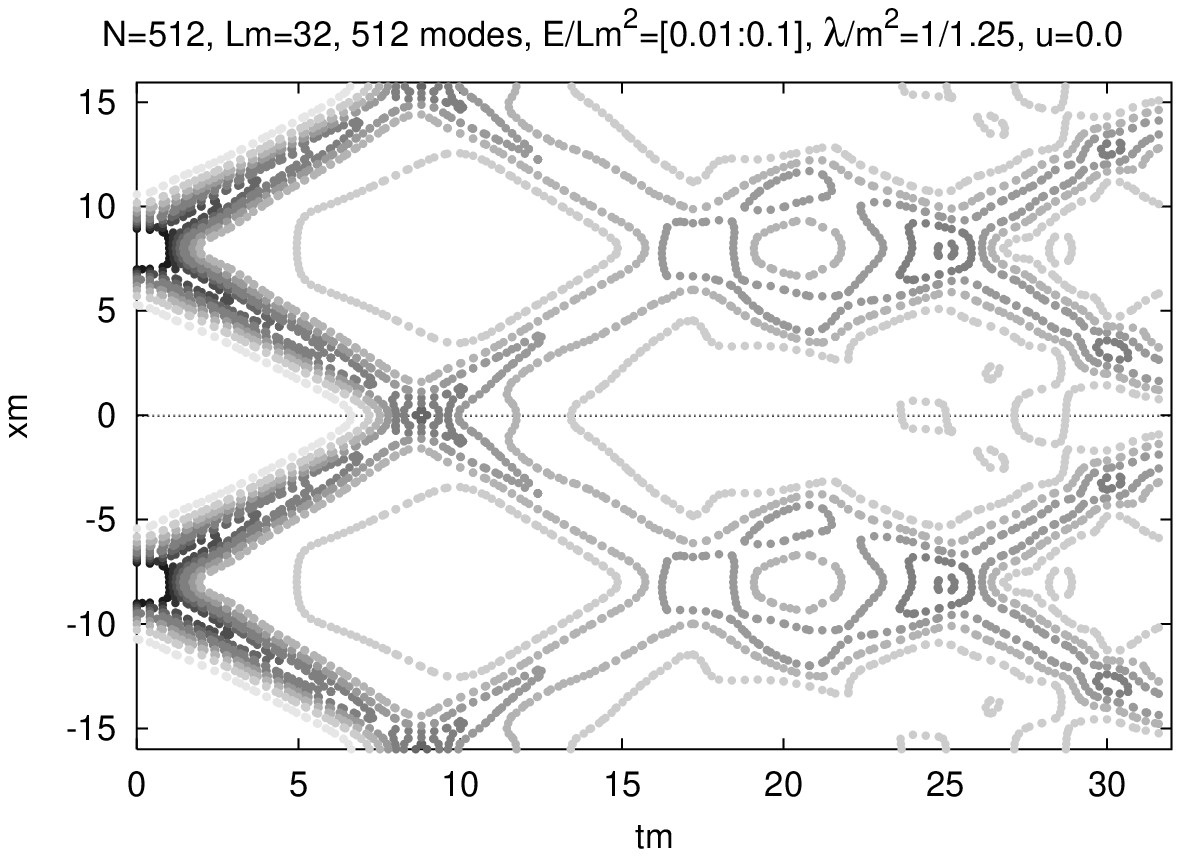}}
\subfigure[Hartree, $\lambda/m^2=1/1.25$, with damping $\Gamma=0.4m$ till
$tm=15$. Plotted energy is mean field only.\label{fig:statHartstrongdamped}]{
    \includegraphics[width=0.475\textwidth]{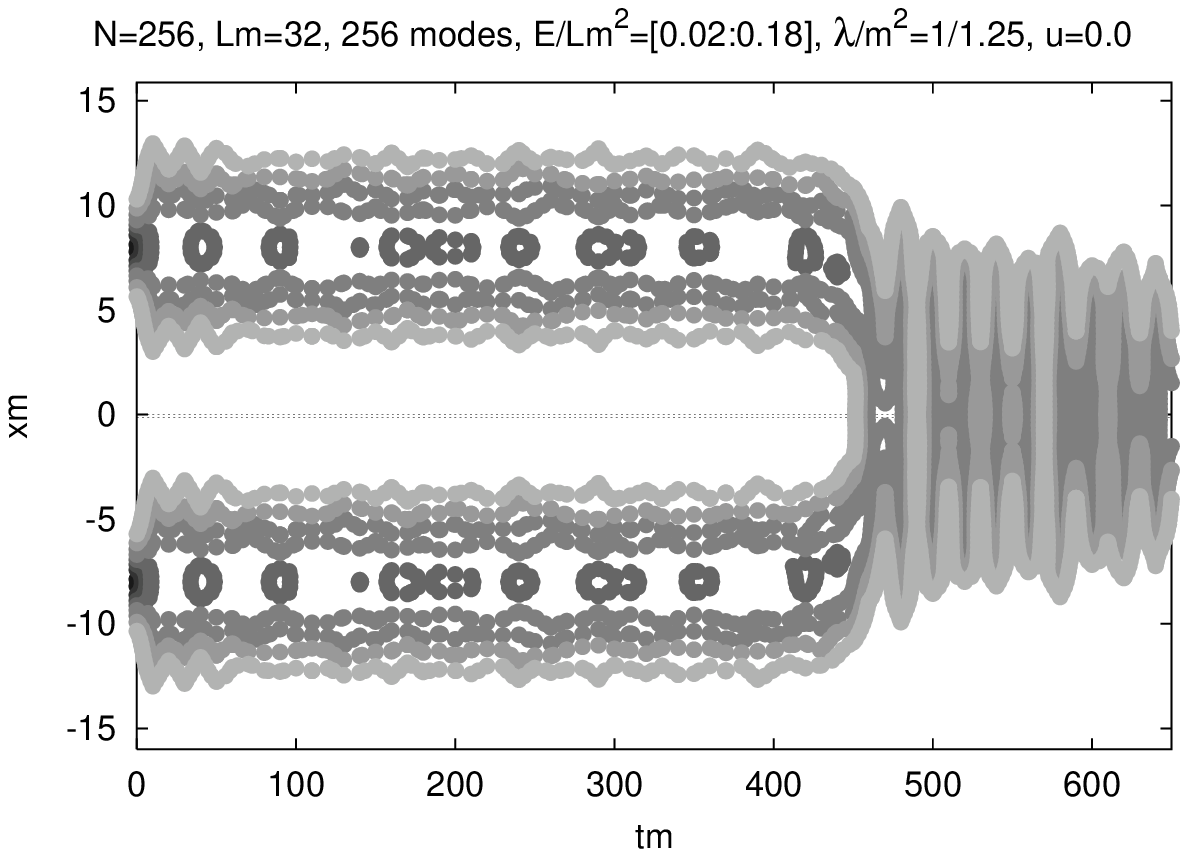}}
\subfigure[Hartree, $\lambda/m^2=1/12$.\label{fig:statHartweak}]{
    \includegraphics[width=0.475\textwidth]{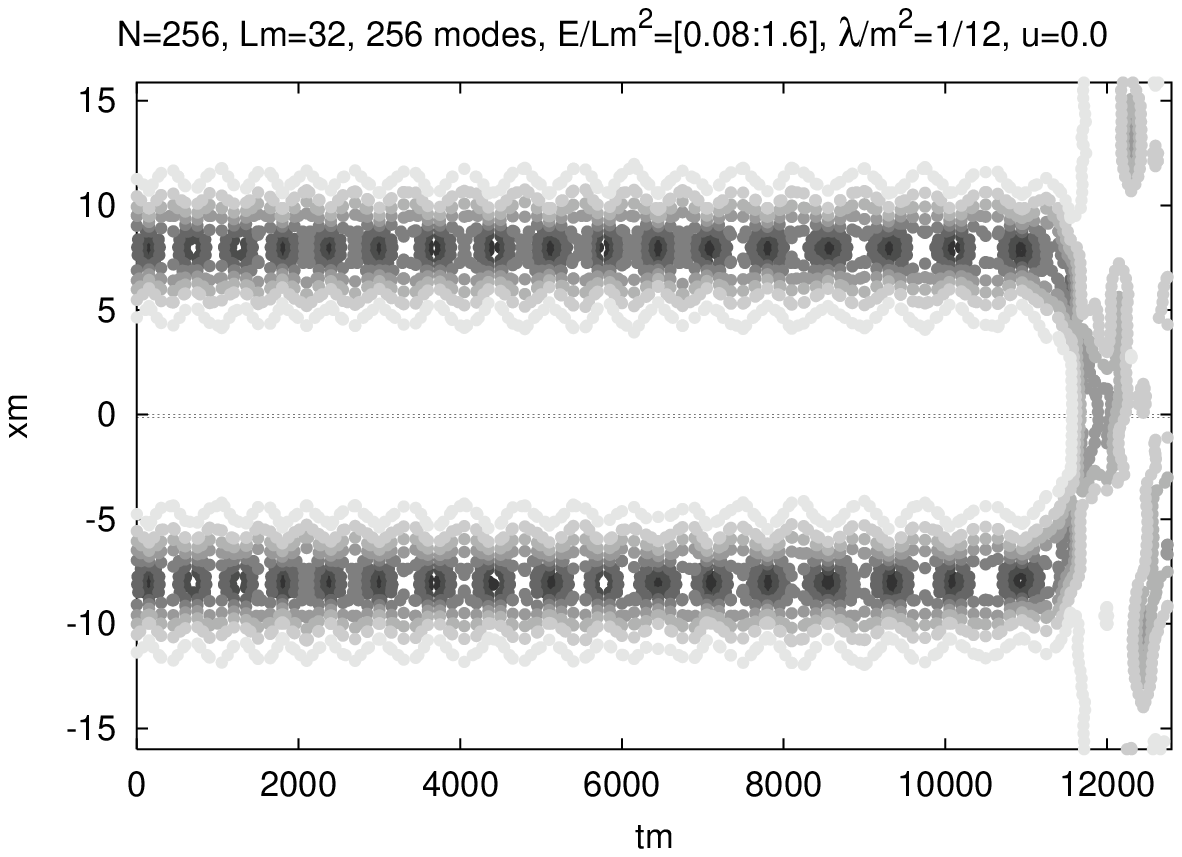}}
\end{center}
\caption[Kink-antikink annihilation from rest.]{Energy density contour plot for
kink-antikink annihilation from rest, for different couplings, classical and
Hartree. All plots, except~\ref{fig:statHartstrongdamped}, show total energy
density.
The vertical energy density range is chosen such as to emphasize the deviation
from vacuum which in some of the plots will cut-off the peak at the center of
the kink. 
\label{fig:statickink}}
\end{figure*}
In this subsection we are discussing the evolution of a kink-antikink
configuration, initially at rest at a maximum distance in the periodic volume. A
comparison will be made between Hartree and classical dynamics, using identical
initial conditions, in the sense that the mean field in the Hartree simulation
is equal to the field in the classical simulation. Hartree simulations at small
and large coupling will also be compared. In all cases the same physical volume
will be used. Finally the effect of damping on the lifetime of the kink-antikink
pair is discussed. The combined results are shown in Fig.~\ref{fig:statickink},
in the form of contour plots of the energy density for the 4 different
simulations. We have made plots of the energy density instead of the (mean)
field, as they show the position of the kink more clearly.
    
Fig.~\ref{fig:statclassstrong} shows a classical simulation at the strong
coupling $\lambda/m^2=1/1.25$. Fig.~\ref{fig:statHartstrong} shows the same
simulation, using Hartree dynamics. Fig.~\ref{fig:statHartstrongdamped} shows
the same Hartree simulation, with a damping $\Gamma=0.4 m$ switched on during
$tm\leq15$. Finally Fig.~\ref{fig:statHartweak} shows the result for a Hartree
simulation at a weaker coupling $\lambda/m^2=1/12$, without a damping term in
the equations of motion.

A comparison of classical and Hartree dynamics, Figs.~\ref{fig:statclassstrong}
and \ref{fig:statHartstrong}, respectively,
shows
an enormous difference in
annihilation time between them. The high instability of the Hartree kink
antikink pair is caused by the fact that at this strong coupling, the average
energy density $e/m^2=0.025$ is higher than the potential barrier $\Delta
e/m^2=0.019$ between the three local minima (remember that the Hartree
approximation incorrectly predicts a first
order phase transition instead of a crossover),
so by exciting particles and spreading the energy the barrier
effectively disappears. The fact that we do not use an actual stationary
solution to the equations of motion \eqref{eq:eom} therefore causes the
$K\bar{K}$ pair to evaporate very quickly. It is important to mention that,
although it seems from Fig.~\ref{fig:statHartstrong} that the kink and antikink
each split into two, this is not actually the case. By making an animation of
the mean field, we found that it just ``contracts'' over the barrier, thereby
annihilating both kink and antikink. In this process some energy is send off in
opposite directions, but this cannot be seen as kink-antikink pairs.

It is also interesting to note that forming a Bose Einstein distribution with
the same average energy density would have an inverse temperature $\beta m=3.5$,
just below the phase transition at which the three minima become degenerate.
Combined with the low barrier this makes it profitable for the configuration to
disintegrate, in order to lower the gradient energy, and settle down in the
symmetric minimum.

Comparing the Hartree results in Figs.~\ref{fig:statHartstrong} and
\ref{fig:statHartstrongdamped} confirms the above interpretation. By damping the
equations of motion we establish two things. We make the configuration closer to
a truly stationary one, thus making it more difficult to excite particles, and
by lowering the average energy density the corresponding Bose Einstein
temperature moves away from the phase transition point.

After the damping is switched off at $tm=15$, the kink-antikink pair survives
until $tm\approx450$. Switching off the damping at a later instant increases the
survival time: using $tm=40$ instead of $tm=15$ results in a survival time of
$tm\approx500$, i.e., the survival time increases by $50$ due to an increase in
``switch-off time'' of $25$. When we do not switch off the damping, the pair
survives until at least $tm=2000$. We have not simulated this system longer, but
we do not think it will ever decay.

We also checked the influence of the precise value of the damping constant
$\Gamma$, by making additional simulations at $\Gamma=0.2m$ and $\Gamma=0.8m$.
At the smaller value, $\Gamma=0.2m$, the pair survival time decreases slightly,
$\Delta tm=-20$, at a larger damping, $\Gamma=0.8m$, the survival time increases
by only $\Delta tm=+2$. Apparently, using a damping constant $\Gamma=0.2m$ until
$tm=15$ already removes most of the fluctuations around the stationary
configuration. However, the remaining configuration is still much less stable
than its classical counterpart. This is
confirmed
by a comparison of the
total energy density with the depicted mean field energy density: it turns out
that the total energy density is much smoother than that of the mean field.
Apparently the modes in part cancel out the kink-antikink
inhomogeneities, thus facilitating
the final annihilation of the pair. 

There is an important subtlety with the implementation of the damping term.
Since it is far from easy to add such a term to the equations of motion of the
mode functions -- even in the vacuum their unrenormalized kinetic term is not
vanishing -- we only implement one in the mean field equation of motion.
However, this makes the actual decrease of the energy density much slower, as it
involves the transfer of energy from the modes to the mean field. We have seen
in the simulations that there are basically two timescales, a rapid initial
decrease, approximately with the damping rate expected, after which most of the
kinetic energy of the mean field has disappeared, followed by a very slow
decrease, draining the energy of the modes. When switching off the damping after
the initial stage, as we have done in the simulation of
Fig.~\ref{fig:statHartstrongdamped}, the kinetic energy of the mean field starts
growing again, be it slowly, and a new energy equilibrium is established. When
the damping is not switched off, the system evolves exponentially slowly towards
its final state. We will come back to this in the next section, in the study of
the Hartree kink mass.

By comparing Fig.~\ref{fig:statHartstrong} at $\lambda/m^2=1/1.25$ with
Fig.~\ref{fig:statHartweak} at $\lambda/m^2=1/12$, we see that at the weaker
coupling, the use of the classical kink form for the mean field and plane waves
for the modes is much closer to a stationary solution, even without damping. The
configuration is oscillating around an approximately stable solution for an
extensive period of time, before annihilation. By comparing
Fig.~\ref{fig:statHartweak} with the damped strong coupling result
Fig.~\ref{fig:statHartstrongdamped}, we see that the kink is broader at the
stronger coupling, due to a change in the effective potential caused by the
quantum modes (we also checked this broadening directly in a plot of the mean
field). Classically the width only depends on the mass $m$, which by
construction is the same for both couplings. This change of width gives another
interpretation for the fact that at larger coupling, the kink-antikink
configuration is less stable: relative to their own size, the separation is
smaller.

As a final remark, it is important to note that the annihilation in
Figs.~\ref{fig:statHartstrong}--\ref{fig:statHartweak} is not due to tunnelling
but is an ``over the barrier'' annihilation. In Fig.~\ref{fig:statclassstrong}
this is clear as it is a purely classical simulation, but one could suspect that
the much shorter decay time in the quantum Hartree simulations is due to
tunnelling. However, in all cases, including the classical, the decay is due to
the attractive force between the kink and the antikink. The attractive potential
falls off exponentially with the separation, and small deviations from the
stationary state only lead to decay after very long times. At strong coupling
the deviations are larger and, as we have seen, in the Hartree simulations the
effective relative distance becomes smaller due to the broader kinks. In all
runs in Figs.~\ref{fig:statickink}, except Fig.~\ref{fig:statHartstrong}, the
kink and antikink are moving towards each other before the decay,
due to this attractive force,
until they bounce and annihilate. After that the mean field ends up in one of
the two ``broken minima''. In Fig.~\ref{fig:statHartstrong}, the barrier is so
low that the field just moves over it and decays in that way. In this run, the
field oscillates around the ``symmetric minimum'' after the decay, as already
discussed before.

\subsection{Numerical results: Kink mass\label{ssec:kinkmass}}

In this subsection we will discuss the kink mass, the energy of the static
solitonic configuration. Classically it is given by
Eq.~\eqref{eq:classkinkmass}. In the full theory this expression needs
corrections. The first corrections were found by Dashen et al.\cite{DaHa74}, see
also Rajaraman~\cite{Ra82},\footnote{Note that there are two errors in
\cite{Ra82}: In Eq.~(5.69) $\sqrt{p^2+1} (p^2+4)$, in the numerator of the first
term in the integral, should be $(p^2+1) \sqrt{p^2+4}$ and $\sqrt{p^2+2}$ in the
numerator of the second term in the integral should be $\sqrt{p^2+4}$. This
second error comes from the transition of $2k^2 \to p^2$.} and more recently by
Alonso Izquierdo et al.\cite{AlGa02}. A numerical study was done in
Ref.~\cite{We99}. The result is obtained using a semiclassical perturbative
expansion around the classical kink solution. The kink mass is then given by the
lowest energy level. This mass has to be renormalized and the net result is
\begin{equation}
M_\text{kink} = \frac{m^3}{3 \lambda} + m \Bigl(\frac{\sqrt{3}}{12} - \frac{3}{2
\pi} \Bigr) + \mathcal{O}(\lambda).
\label{eq:kinkmass1loop}
\end{equation}
A partial calculation of the order $\lambda$ correction can be
found in Ref.~\cite{Ve76}. In this reference and independently also in
Ref.~\cite{Ve77} a full calculation
is done for the
order $\lambda$ correction of the
kink mass in the sine-Gordon model.

More recently lattice Monte Carlo studies of the kink mass have been carried out
\cite{CiTa94,ArWi99}. In these papers two methods of calculating the kink mass
were used: first using the fluctuation-fluctuation 2-point function
$\braket{\mu(t) \mu(0)}$ introduced in Refs.~\cite{Ka69,KaCe71}, which in
imaginary time should decay as $\exp(-M_\text{kink} t)$. The second uses the
difference in the ground state energy when periodic or antiperiodic boundary
conditions are used. Both studies focus on a range of parameters, where $\lambda
a^2$ and $-\mu_0^2 a^2$ are of order unity, i.e., at large lattice
distances.
\begin{figure}[tbp]
\begin{center}
\subfigure[$a^2 \mu^2=-1.0$.\label{fig:kinkmass_mu1}]{
    \includegraphics[width=0.4575\textwidth]{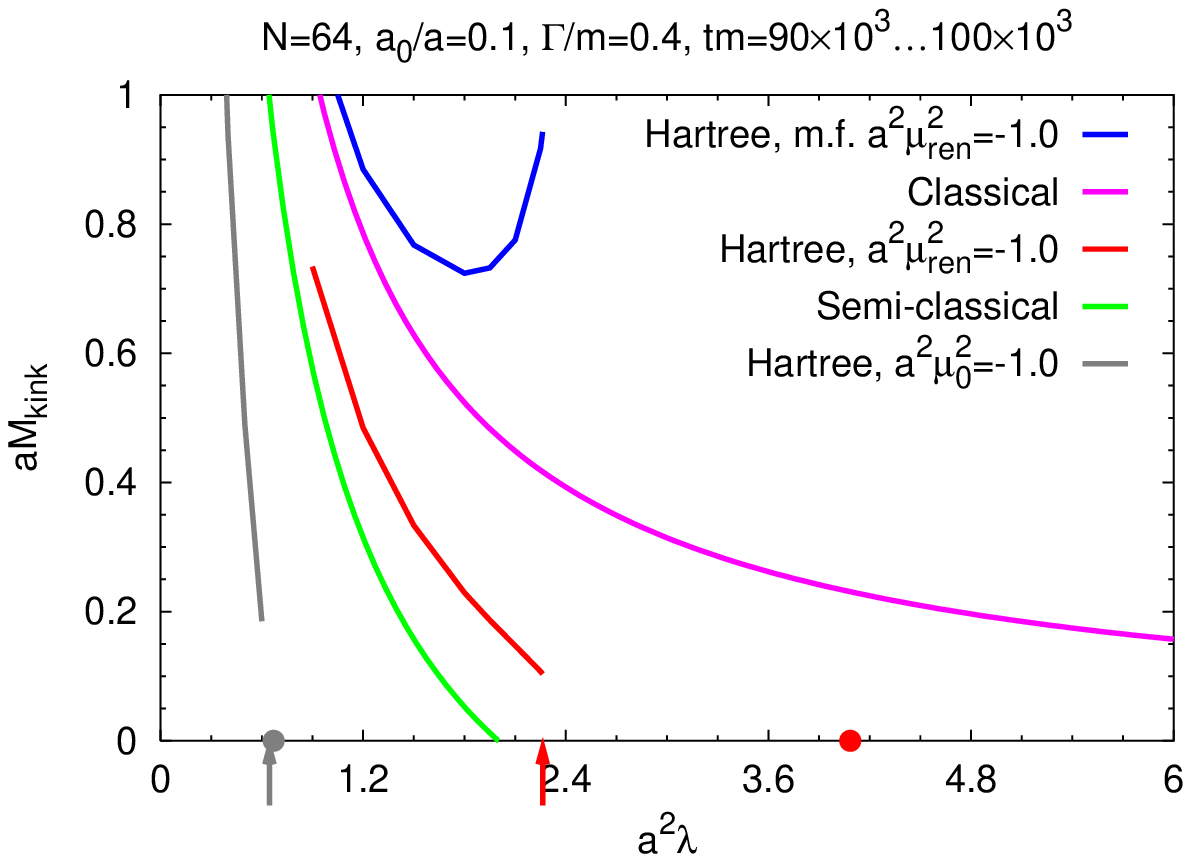}
}
\subfigure[$a^2 \mu^2=-2.2$.]{
    \includegraphics[width=0.4575\textwidth]{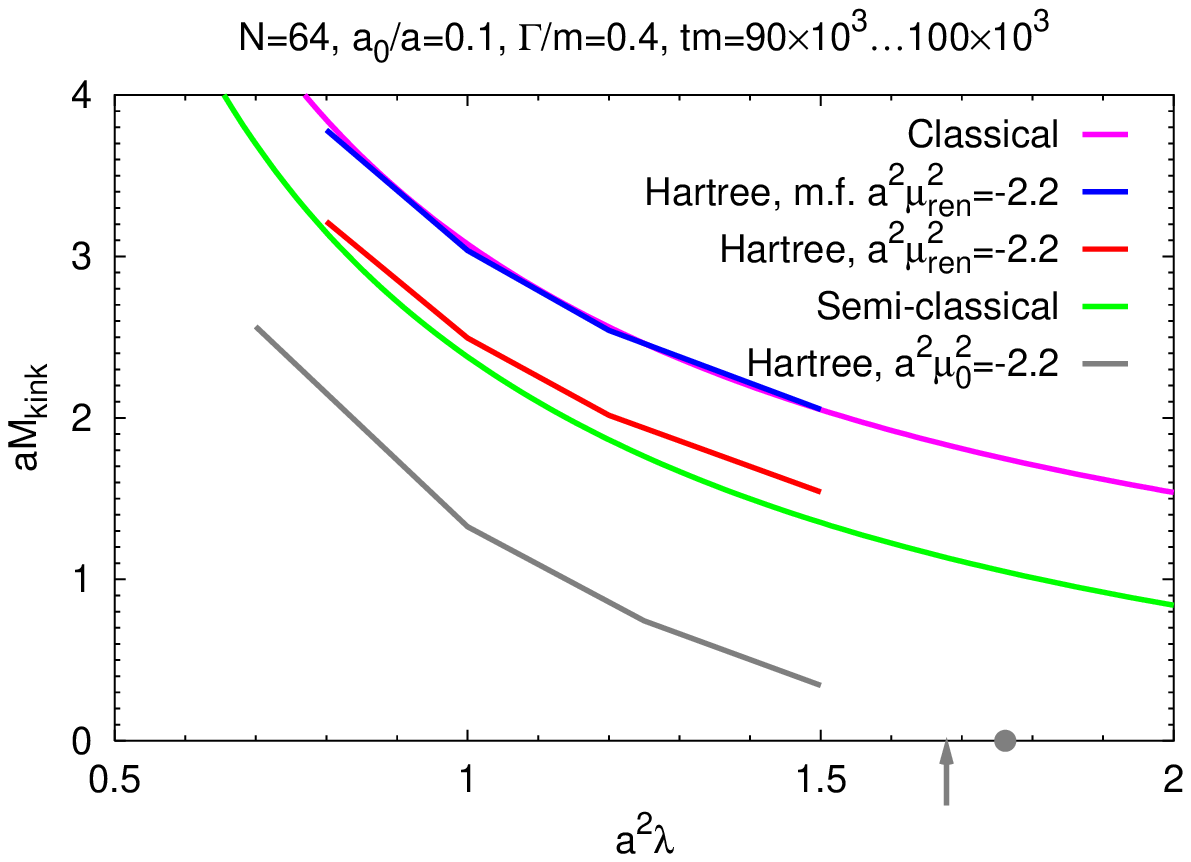}
}
\subfigure[$a^2 \mu^2=-4.0$.]{
    \includegraphics[width=0.4575\textwidth]{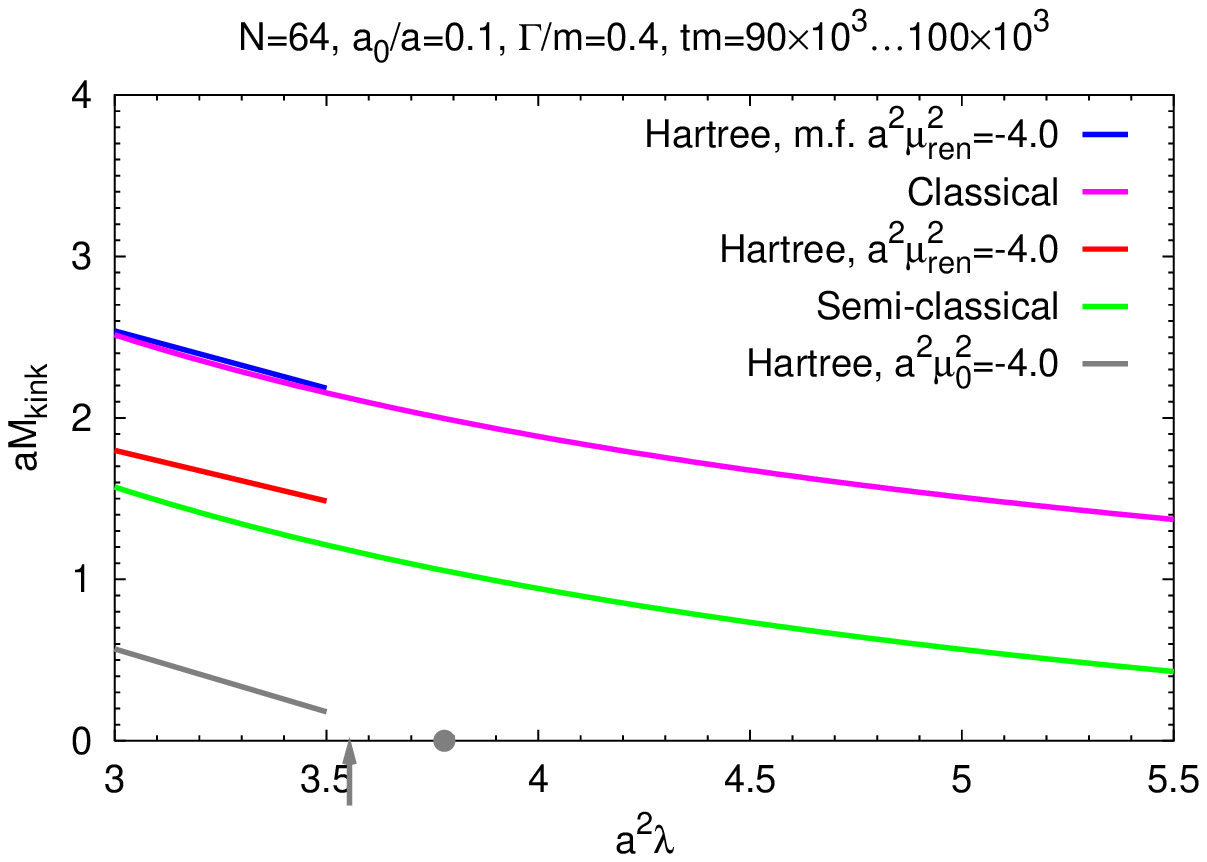}
}
\end{center}
\caption{Comparison of classical, semiclassical (one loop) and Hartree
results. The Hartree results are plotted for constant renormalized
$\mu_{\text{ren}}^2$ and unrenormalized $\mu_0^2$. Also see the
text.\label{fig:kinkmass}}
\end{figure}

Very recently, Bergner and Bettencourt~\cite{BeBe03} studied the kink mass in
the Hartree approximation. Their equations are derived from a 2PI effective
action, but applying the Hartree approximation to it results in the same
equations of motion we are considering here. Using a numerical relaxation method
they have obtained the Hartree kink mass at $\lambda/m^2=0.5$ to be $0.4691$
(translated in our units), which is about $70\%$ of the classical value. We have
reproduced their value using antiperiodic boundary conditions and a damping term
\eqref{eq:dampterm} in the mean field equation of motion. On a lattice with
$N=64$ and $Lm=16$, $a/a_0=10$ using a damping coefficient $\Gamma/m=0.4$ we
find at a time $tm = 9 \times 10^3 - 10 \times 10^3$ a value
$M_\text{kink}/m=0.47$. Doubling the lattice size or decreasing the temporal
lattice distance increases
this
value by about $1\%$,
while
increasing the simulation
time with an order of magnitude decreases it by about $2\%$. Unlike Bergner and
Bettencourt we have not fixed the center of the kink, but the fact that the mass
turns out equal shows that the zero mode can have at most a small effect,
in agreement
with Ref.~\cite{ArWi99}.

In order to make a detailed comparison with the full theory, i.e., the Monte
Carlo data of Refs.~\cite{CiTa94,ArWi99}\footnote{Note that the horizontal axis,
$\hat{\lambda}$, in Fig.~4 of Ref.~\cite{ArWi99} should be multiplied by $6$, as
can be derived from the curve of the semiclassical kink mass.}, we have done
simulations for a range of parameters. The results are shown in
Fig.~\ref{fig:kinkmass}. We show results for constant renormalized
$\mu_\text{ren}^2$ as well as constant unrenormalized $\mu_0^2$.
For the constant $\mu_\text{ren}^2$ we also show separately the
mean field contribution, defined as in Eq.~(16), Ref.~\cite{SaSm02}.
The data in
Refs.~\cite{CiTa94,ArWi99} is shown as a function of
the unrenormalized $\mu_0^2$,
in order to
study the renormalization effects. However, to make a meaningful comparison with
the classical and semiclassical 1-loop results, it seems more proper to compare
at constant $\mu_\text{ren}^2$. The arrows in the plots show the zero
temperature phase transition point, at which the three minima become degenerate,
calculated using the static effective potential, cf. Ref.~\cite{SaSm00a}. The
dots indicate the point at which the ``broken'' minima vanish altogether. In
order to make the comparison with the Monte Carlo data easier, lattice units
have been used.
Most
simulations where done with $N=64$, $a_0/a=0.1$ with a damping term
$\Gamma/m=0.4$ while the energy was measured at $tm\approx100\times10^3$. We
have checked for lattice artefacts (the influence of $N$ and $a_0/a$), their
combined error is not larger than about $1\%$. Also the effects of the precise
value of the damping term and time at which the mass is measured are small and
of the same order of magnitude as that of lattice artefacts.

We see that for all renormalized masses, the Hartree results are between the
classical and semiclassical values. Using the unrenormalized mass as the
parameter to compare with, the Hartree results are much lower than the
semiclassical values. Comparing these with Figs.~4--6 of Ref.~\cite{ArWi99} we
find that the Monte Carlo results are above the Hartree results. Note that the
semiclassical line in Fig.~6 of
Ref.~\cite{ArWi99}
is approximately $0.3$ too high.
For the two higher values of
$|\mu_\text{ren}^2|$,
where the data taken is further
away from the first order phase transition, the mean field contribution is
almost equal to the classical value. The difference between the Hartree and
classical kink mass only comes from the modes. For the lower
$a^2 |\mu_\text{ren}^2|=1.0$
the ``mean field mass'' behaves very interestingly. We
see that, coming from below, it goes up again near the phase transition. This is
caused by the emergence of the symmetric minimum. More data would be needed to
study the precise behaviour at the phase transition, but from its definition,
Eq.~(16a) in Ref.~\cite{SaSm02}, it seems it only goes up a final amount, due
to this symmetric minimum and the change in the effective potential. It is also
not clear to us at this stage whether it is only an artefact of the way we split
the total energy in a mean field and mode contribution or is an actual physical
phenomenon.

Finally, we have measured the mass at a coupling $\lambda/m^2=1/1.25$, used for
example in Figs.~\ref{fig:statHartstrong} and \ref{fig:statHartstrongdamped},
using $Lm=16$, $N=64$ at a time $tm=20\times10^3$. In principle we could find
the value also in Fig.~\ref{fig:kinkmass_mu1}, but those data where obtained a
large lattice distance and a separate test is useful. The result is
$M_\text{kink}/m=0.22$, or $54\%$ of the classical value
$0.42$. This
value for
the kink mass
is consistent with Fig.~\ref{fig:kinkmass_mu1}, showing
that we are indeed in the scaling region, as claimed by Ref.~\cite{ArWi99} (the
same is also found for the weaker coupling $\lambda/m^2=0.5$, considered in
Ref.~\cite{BeBe03}). The semiclassical value \eqref{eq:kinkmass1loop} from
Ref.~\cite{DaHa74} gives a value $0.08$, about $1/5$ of the classical value.
Note that at $\lambda/m^2 \approx 1$ the semiclassical approximation breaks
down, showing up by a change of sign in the kink mass, the order $\lambda^0$
correction becomes equal to the leading $\lambda^{-1}$ term. In the Hartree
approximation we can in principle continue till $\lambda/m^2 \approx 1.2$, where
the zero temperature phase transition occurs.

\section{Moving Hartree kinks\label{sec:coll_kink}}

\subsection{Initial condition}

In order to describe kink-antikink collisions, it is necessary to have a
description of a moving Hartree kink as well. In the full quantum theory this
poses a difficult problem, since we do not know how to
Lorentz
transform the ``quantum
cloud'' surrounding the kink. In the Hartree approximation we are saved by the
fact that the field is completely expressed in terms of ordinary functions of
$x$ and $t$, and we can find a moving kink by performing a simple Lorentz
transformation:
\begin{equation}
f(x,t) \to f'(x,t) = f(\gamma [x-u t],\gamma[t-u x]),
\label{eq:lorentz}
\end{equation}
both on the mode functions and the mean field. 

However, there are still a number of difficulties with finding a proper initial
condition for a kink-antikink pair, moving toward each other. First of all, from
Eq.~\eqref{eq:lorentz} it follows that we need the mode functions on a backwards
space-time line, while we do not know them in analytic form, only from numerical
simulations. For the classical kink solution this problem does not exist, since
it is not only stationary, but also static and known analytically. Second, we
have to boost the kink and antikink, with their respective mode-functions, in
opposite directions. This means that they have to be combined in a nontrivial
way in the middle, they ``shift into each other''. It also means that we have to
double the density of mode functions in $k$-space, since the physical space
doubles (the best one can do is determine the mode functions for a single kink).
We have to decide how to do this in a consistent way. Finally there is the
problem that the boosted coordinates $\gamma(x-u t)$ and $\gamma(t-u x)$ will in
general not fall onto space-time lattice points in the unboosted frame or vice
versa.

In order to circumvent these problems, we will just use the vacuum form as
initial conditions for the mode functions. Since the vacuum is invariant under
Lorentz transformations, we can use the same mode functions in a boosted frame.
As mentioned before, the mode functions will be close to the vacuum form, if the
vacuum expectation value $v$ is large.
Furthermore by
boosting the kink solutions we reduce
the width of the kinks. However, in the center of mass frame, we keep the
lattice size the same thus effectively increasing the relative distance between
them. This leads to much more stable solutions,  which is even further enhanced
by the time dilatation. 

Before we proceed we will outline a set of solutions for the aforementioned
problems, which will yield a very accurate, boosted, Hartree kink-antikink
solution and will therefore also be usable at strong coupling and low speeds. In
the simulations we will in general, for simplicity, only use the vacuum form
mode functions.

The way to solve the stability problem of a (single) kink is to use damped
equations of motion, as we already used in the preceding subsections, especially
in combination with antiperiodic boundary conditions on the mean field to make
the kink absolutely stable. The damping will give us stationary, but
time-dependent solutions for the mode functions in the background of a kink,
which subsequently can be Lorentz boosted.

The problem of the boosted coordinates not resulting from discrete lattice
points in the unboosted coordinate can be adequately solved by linear
interpolation, provided the lattice distance is not too large. The related
matching problem is not a real problem: we have to match the two functions at
two consecutive time steps and when they approach each other with speed $u$, they
shift into each other over a distance
$u a_0 \ll a$,
i.e., the mismatch is less
than a lattice distance and the linear interpolation automatically solves the
problem. 

The remaining problem is to combine the two sets of mode functions, of the
separate kink and antikink configurations, into one set which is twice as large.
One possible solution is to determine both kink and antikink in a volume
that already has the size of the combined configuration and then use an
averaging procedure to combine the modes. This has the disadvantage that
oscillatory functions with different phases are added. Another way, which we
suggest, is to determine the two configurations in their original volume and
combine them afterwards into one set in the double volume. When using periodic
mode functions, it follows from the antiperiodicity of the mean field that at
all times
\begin{equation}
f_k(x,t) = f_{-k}(L-x,t),
\end{equation}
provided this relation holds at $t=0$ and $t=a_0$, a condition which is true for
the plane wave initial conditions we use. Note that at later times the $k$ label
is no longer equal to the Fourier label, for which the relation holds trivially.
We can therefore combine each mode function $f_k$ with either itself or with
$f_{-k}$, since both combinations yield continuous functions. The derivative
will in general be discontinuous by changing sign, possibly causing trouble in
the equation of motion. However, the energy density will be finite and even
continuous, as it only depends on the square of the first derivative, which is
continuous. 

We end this subsection summarizing the above steps one could follow to obtain a
better initial state for a colliding kink-antikink pair.
First, using antiperiodic boundary conditions for the mean field and a damping
term $-\Gamma \partial_t \phi$, one evolves the Hartree equations, starting from
a classical kink with free field mode functions. One thus obtains a stationary
Hartree solution for a single kink, but has to keep track of the modes and mean
field on a backward space-time line. Then one constructs the mean field from the
kink and its mirror image. The modes are constructed by combining $f_k(x,t)$
with itself and with $f_{-k}(x,t)=f_k(L-x,t)$, thereby doubling their number,
\begin{subequations}
\begin{align}
f_{k,1}(x,t)&=\begin{cases}
f_k(x,t), & 0 \le x \le L, \\
f_k(x-L,t), & L \le x \le 2L,
\end{cases} \\
f_{k,2}(x,t)&=\begin{cases}
f_k(x,t), & 0 \le x \le L, \\
f_{-k}(x-L,t), & L \le x \le 2L.
\end{cases}
\end{align}
\end{subequations}
The
latter, symmetric,
combination can lead to functions in which the second derivative
diverges as $1/a$: the energy density, depending only on the
square of the
first
derivative,
is
finite and continuous in the continuum limit, however it might cause problems in
the equations of motion.

\subsection{Numerical results: Thermalization from colliding kink-antikinks}

We start the discussion of our numerical results with the thermalization
properties of a colliding kink-antikink pair, using the defining relation
\eqref{eq:studef} for the particle number $n_k$. We used a small coupling,
$\lambda/m^2=1/12$, in a volume $Lm=32$ at a speed for which $\gamma=2$. The
classical kink mass at this coupling is $M_{\text{kink}}/m=4$. Since we
initialize the mean field as a classical kink and the modes in the vacuum, the
initial energy density is $E/Lm^2=2\cdot(4+4)/32=0.5$. At this energy density
and coupling we found, in
Ref.~\cite{SaSm00a},
both in dynamical simulations and
an effective potential calculation, a temperature $T/m\approx1.0$.

\begin{figure}[tbp]
\includegraphics[width=0.483\textwidth]{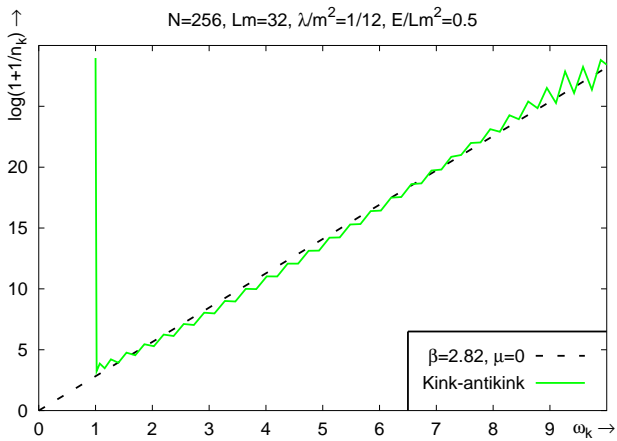}
\caption{Initial form ($tm=0$) of the particle distribution for a Hartree
kink-antikink collision. The initial form already seems close to a Bose-Einstein
distribution.\label{fig:kink_thermal_start}}
\end{figure}

In order to enable us to correctly interpret the result, we also measured the
``particle number'' at $tm=0$. Of course at this time the system is so far from
equilibrium that we cannot interpret the result as quasiparticles, but it does
give an idea of the energy distribution as a function of $k$. The result is
plotted in Fig.~\ref{fig:kink_thermal_start}. The similarity with a
Bose-Einstein distribution is remarkable and we have to be careful in
interpreting the results at later times. One of the indications, that it is not
a truly thermal distribution, is its very low ``temperature'': at an energy
density $E/Lm^2=0.5$ we expect to find a temperature around $T/m=1.0$, while
Fig.~\ref{fig:kink_thermal_start} seems to indicate a value which is almost a
factor of 3 smaller. Furthermore, the particle number mainly comes from the
even modes, as a result of the symmetric initial condition.

\begin{figure}[tbp]
\includegraphics[width=0.483\textwidth]{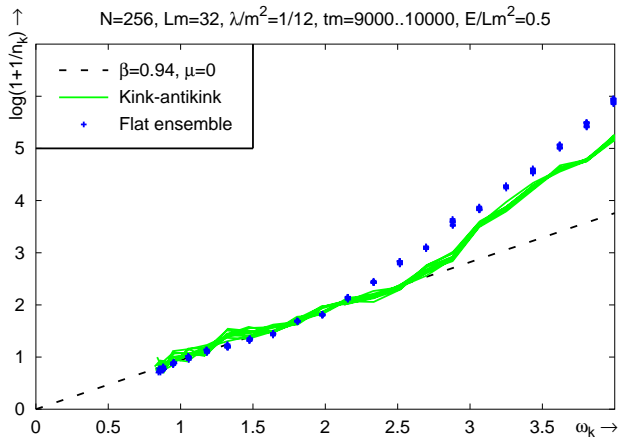}
\caption{Late time form of the particle distribution for a Hartree kink-antikink
collision. $tm\approx10000$, after annihilation.\label{fig:kink_thermal_therm}}
\end{figure}

Fig.~\ref{fig:kink_thermal_therm} shows the actual particle number, at later
times, when the kink-antikink have already annihilated. Then, we do find a
temperature around $1$, conform the results of Ref.~\cite{SaSm00a}. As a
comparison we also plotted, in the same figure, the result obtained in that
paper using a flat ensemble simulation. It clearly shows that the temperatures
are equal: $T$ is uniquely determined by the coupling and energy, not by the
initial condition. The initial condition still has an influence in determining
which modes have thermalized. We see that starting from an annihilating
$K\overline{K}$ actually allows the system to thermalize faster, as it has a
more favourable energy distribution, cf. Fig~\ref{fig:kink_thermal_start}.
However, the $K\overline{K}$ initial condition has a delta function initial
density matrix, which is not very suitable for our Hartree ensemble
approximation: we do not reduce the statistical errors by averaging over
multiple initial conditions.

Having showed that this initial condition also leads to an approximately
thermal Bose-Einstein distribution after annihilation, we will now leave the
thermalization topic and continue with a discussion of the actual collisions.

\subsection{Numerical results: Critical speed\label{ssec:critspeed}}

We start this section with a study of \emph{classical} kink-antikink collisions.
An extensive study of this can be found in Ref.~\cite{CaSc83}.

\begin{figure*}[tbp]
\begin{center}
\subfigure[$u_i=0.19$ ($\gamma\approx1.019$).]{
    \includegraphics[width=0.475\textwidth]
    {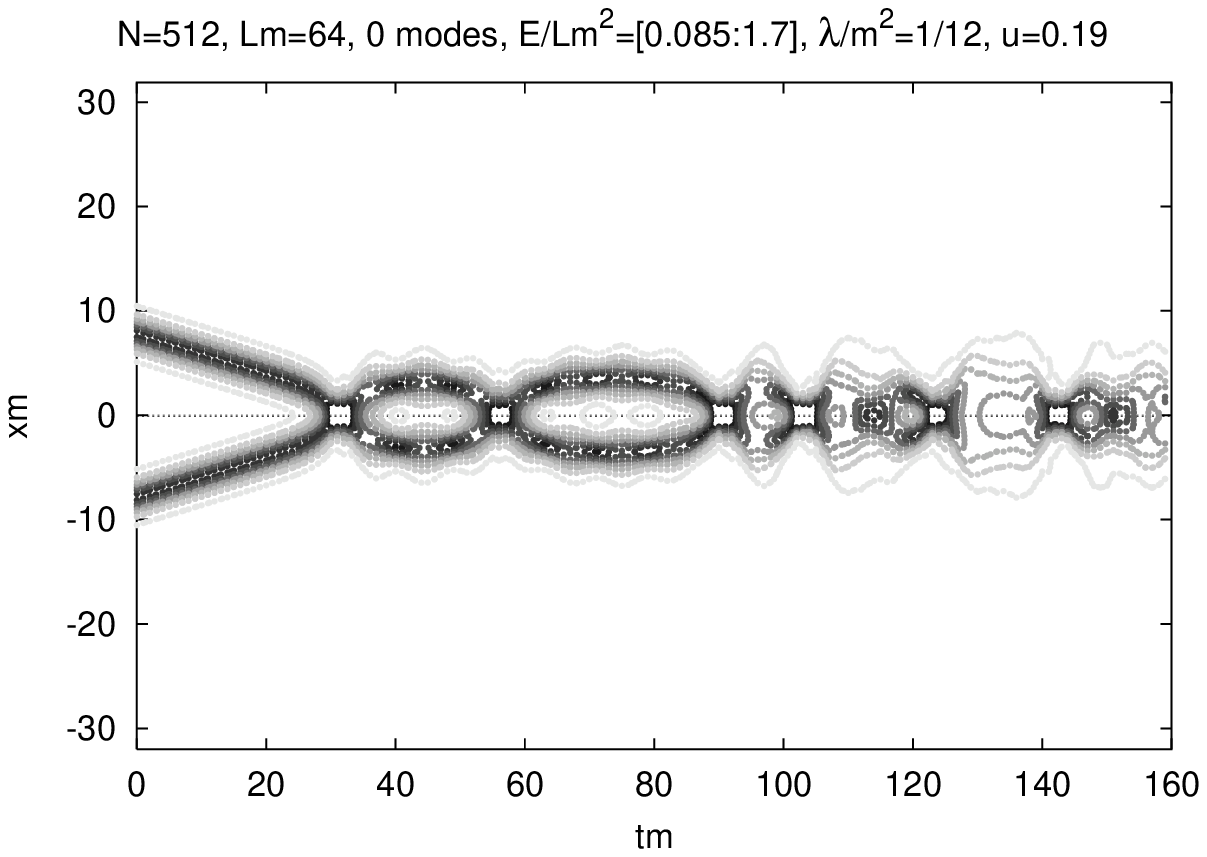}}
\subfigure[$u_i=0.20$ ($\gamma\approx1.021$).]{
    \includegraphics[width=0.475\textwidth]
    {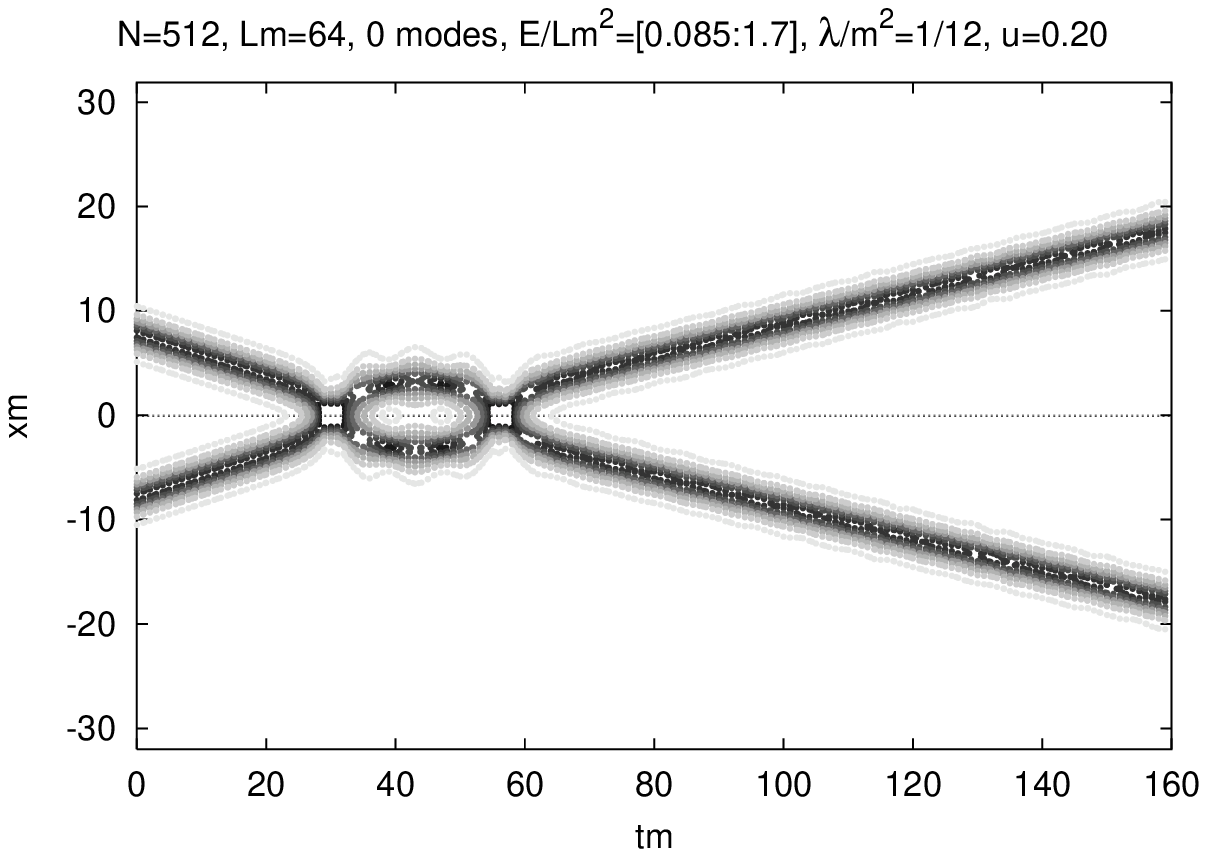}}
\subfigure[$u_i=0.21$ ($\gamma\approx1.023$).]{
    \includegraphics[width=0.475\textwidth]
    {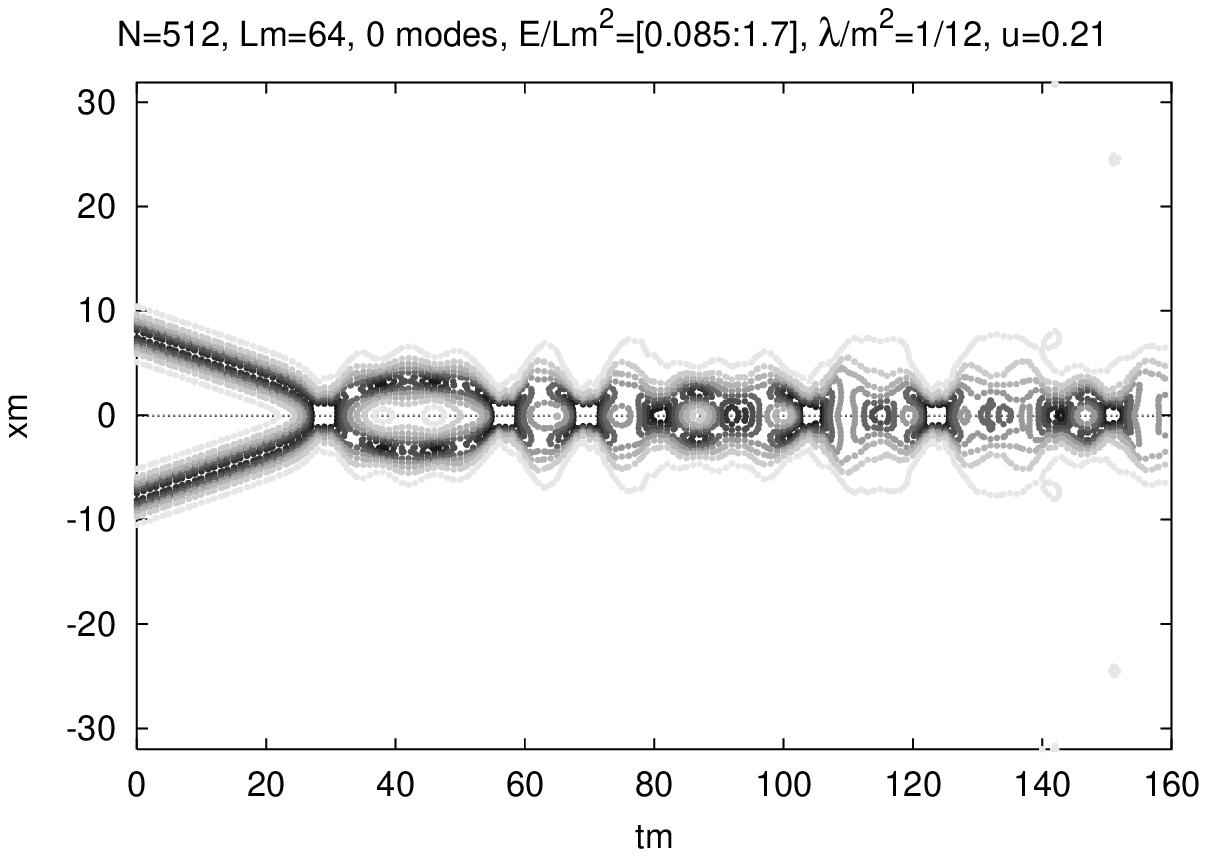}}
\subfigure[$u_i=0.25$ ($\gamma\approx1.033$).\label{fig:mismatch}]{
    \includegraphics[width=0.475\textwidth]
    {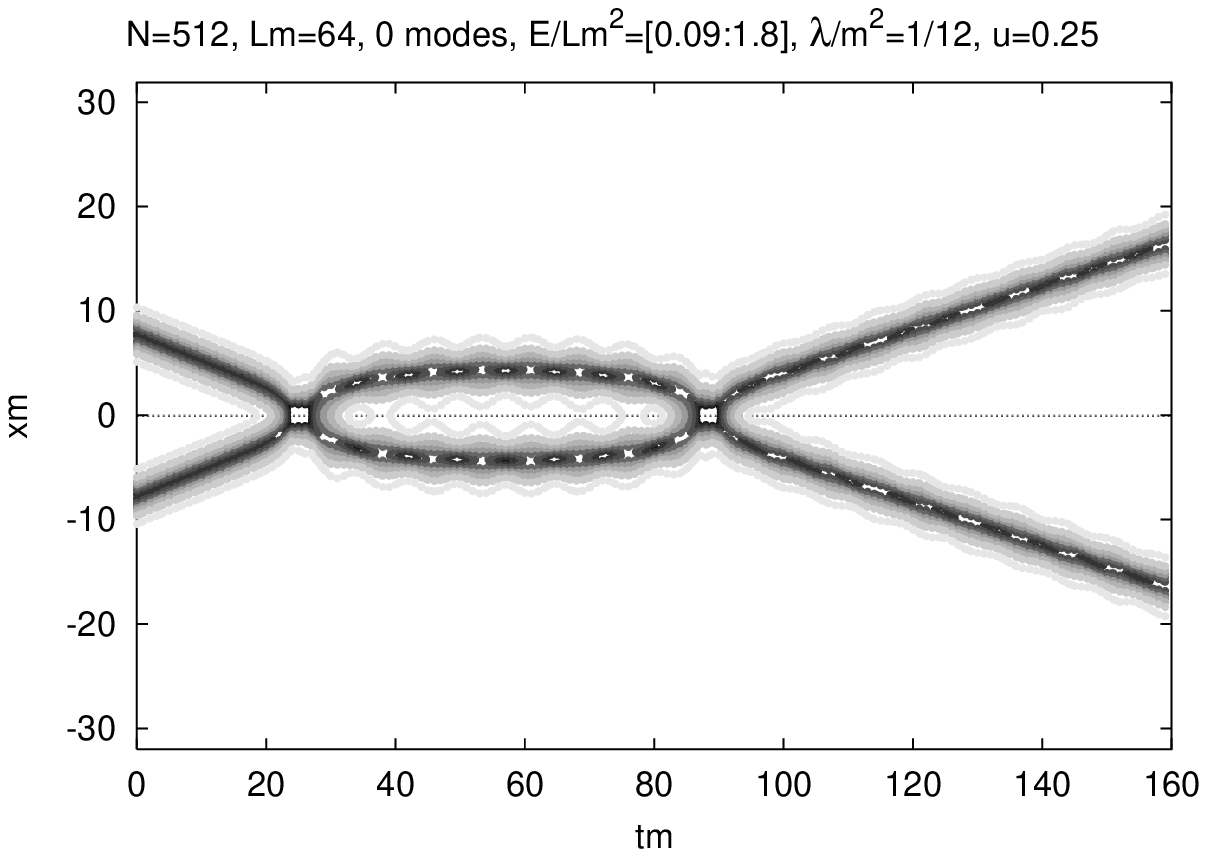}}
\subfigure[$u_i=0.30$ ($\gamma\approx1.048$).\label{fig:abovecrit}]{
    \includegraphics[width=0.475\textwidth]
    {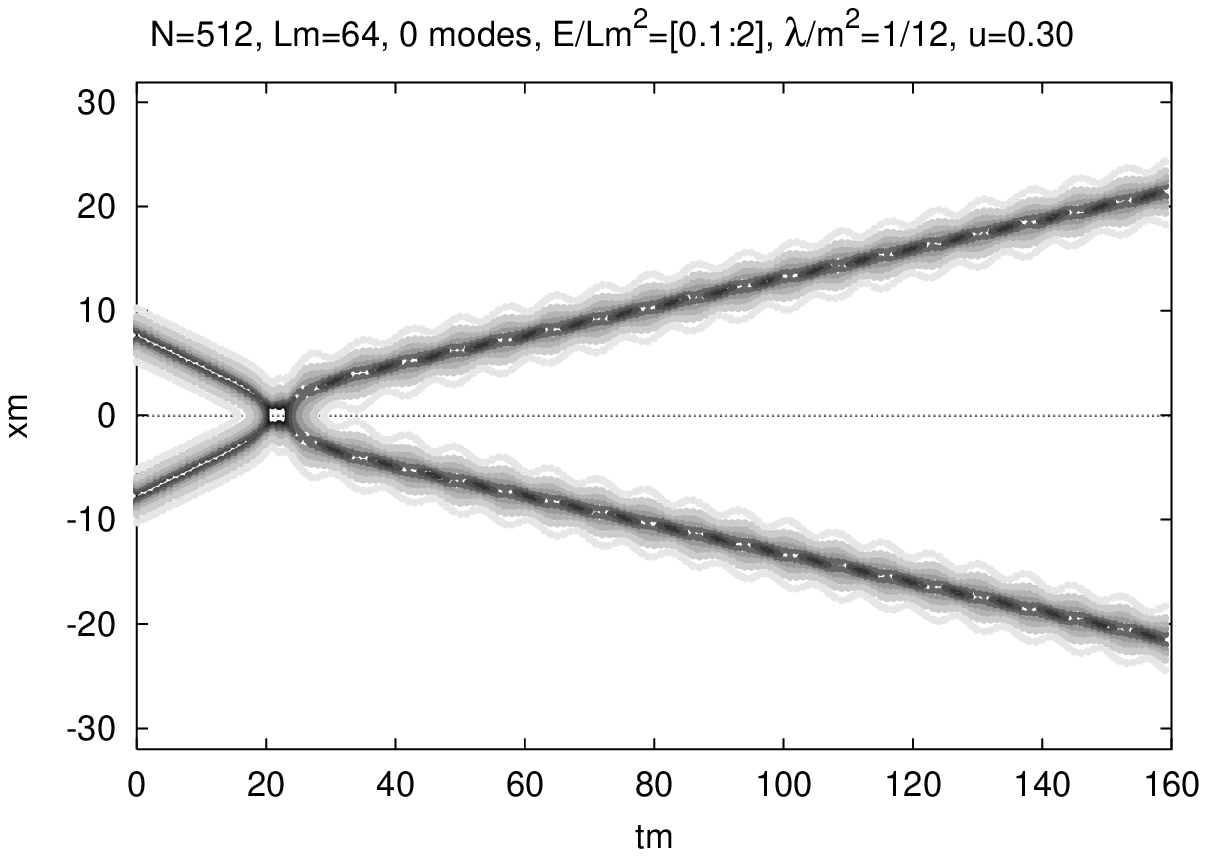}}
\end{center}
\caption[Colliding classical kink-antikink]{Colliding classical
kink-antikink, $\lambda/m^2=1/12$, above and below the critical
speed.\label{fig:kinkcritclass}}
\end{figure*}

Figs.~\ref{fig:kinkcritclass} show energy contour plots of Lorentz boosted
classical kink-antikink configurations. The results in these plots demonstrate
how, at very low incident speed, they just annihilate, while at a certain range
of higher speeds, they bounce a few times and then either annihilate or escape
again to infinity, at a slightly lower speed. At high enough speeds, when $u$
is larger than a certain $u_c$,
which has
a value between $0.25$ and $0.30$, the pair
immediately escapes after the collision, again with a lower speed. Part of the
kinetic energy is transferred to an internal vibrational mode of the kink, which
can be seen as a wiggling in the contour plots.

These results are consistent with Ref.~\cite{CaSc83} in which a critical speed
$u_c=0.2598$ is found, above which colliding kinks always bounce back, while
below a speed $0.193$, they always annihilate immediately. Between these speeds
Ref.~\cite{CaSc83}
found reflection and annihilation bands, caused by a resonance between the
center of mass motion and the vibrational mode. This was confirmed in
Ref.~\cite{AnOl91}, also showing that, when zooming in on the bands, their
structure behaves fractally. This behaviour could even be reproduced using a
collective coordinates approach, reducing the infinite-dimensional system to a
two dimensional Hamiltonian one.

Except for Fig.~\ref{fig:mismatch} all shown results are in accordance with
Table~I in Ref.~\cite{CaSc83}. Since we used different discretizations, etc.,
and
given the very narrow width of the stability bands already found by
Ref.~\cite{CaSc83} and the fractal behaviour found in Ref.~\cite{AnOl91}, we can
certainly expect small differences in
the precise position of these bands,
explaining the
discrepancy. The authors of Ref.~\cite{CaSc83} found that the final speed, for
an initial speed above $u_c$, satisfies the following relation
\begin{equation}
u_f^2 \propto u_i^2-u_c^2,
\label{eq:critspeed_class}
\end{equation}
which is consistent with Fig.~\ref{fig:abovecrit}, from which we derive a final
speed $u_f\approx0.13-0.14$. For $u_i=0.30$ relation \eqref{eq:critspeed_class}
predicts $u_f=0.15$.

It is important to note that although breatherlike states seem to emerge after
a collision, truly stable breathers do not exist in the $\lambda \phi^4$ theory
\cite{SeKr87}. However, very long-lived and almost stationary configurations do
exist \cite{Ge94} as we have shown here again. The radiation causes the energy
to decay as $1/\log(t)$, as found analytically by Ref.~\cite{SeKr87} and
confirmed numerically by Ref.~\cite{Ge94}. 

We would like to compare these classical results with the Hartree approximation
to see if, in the quantum theory, a critical speed still exists. In
the classical theory, one can always use units such that $\mu=\lambda=1$, and
one expects a unique critical speed, while in the quantum theory, including the
Hartree approximation, this is not the case. However, we will not go much
further into the question of the coupling dependence of the critical speed.

\begin{figure*}[tbp]
\begin{center}
\subfigure[$u_i=0.700$ ($\gamma\approx1.400$).]{
    \includegraphics[width=0.475\textwidth]
    {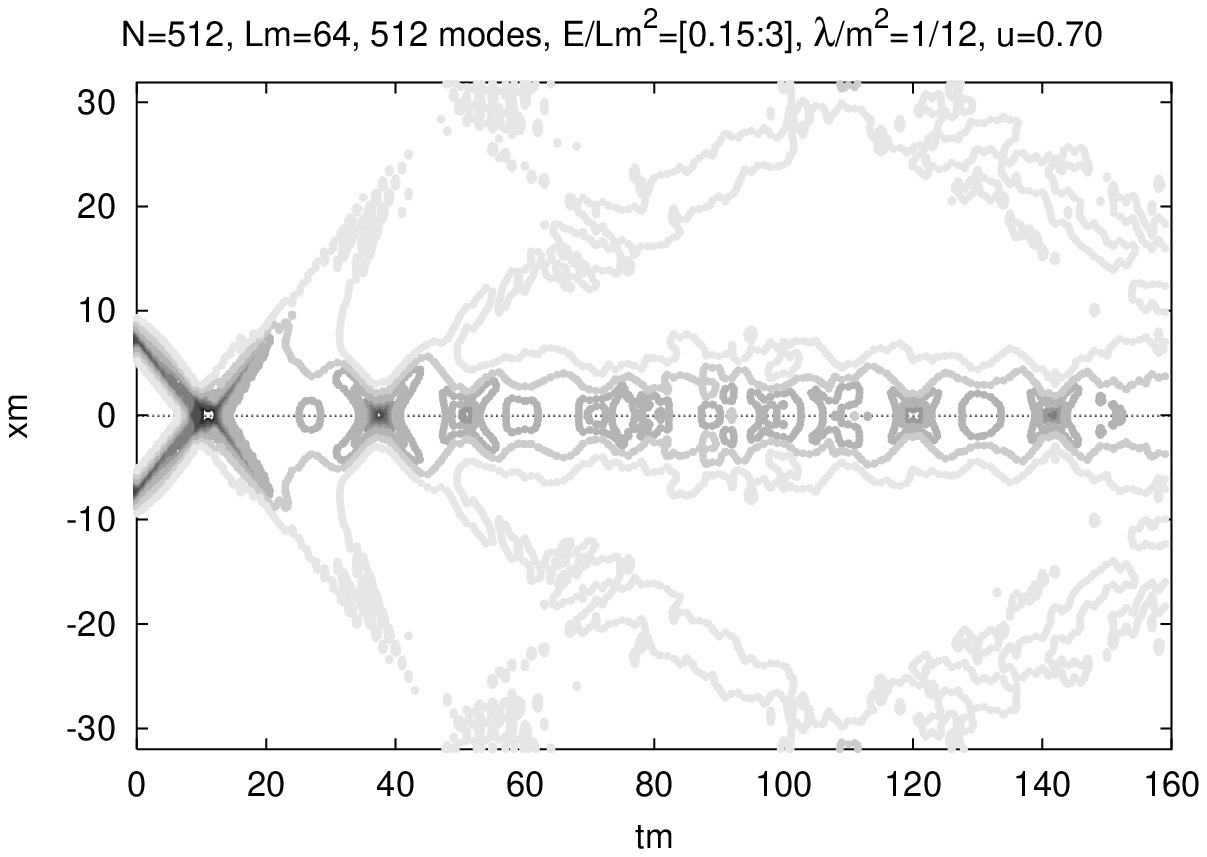}}
\subfigure[$u_i=0.750$ ($\gamma\approx1.512$).]{
    \includegraphics[width=0.475\textwidth]
    {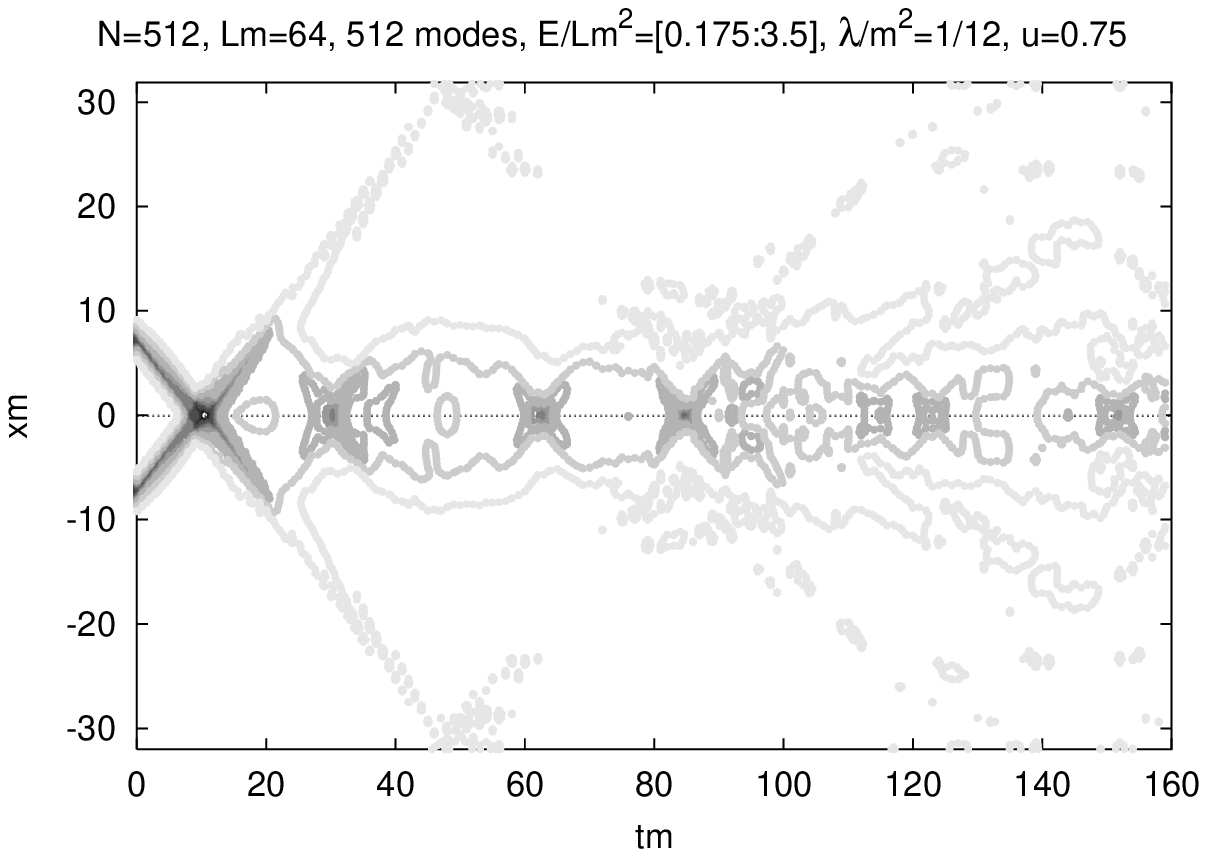}}
\subfigure[$u_i=0.760$ ($\gamma\approx1.539$).]{
    \includegraphics[width=0.475\textwidth]
    {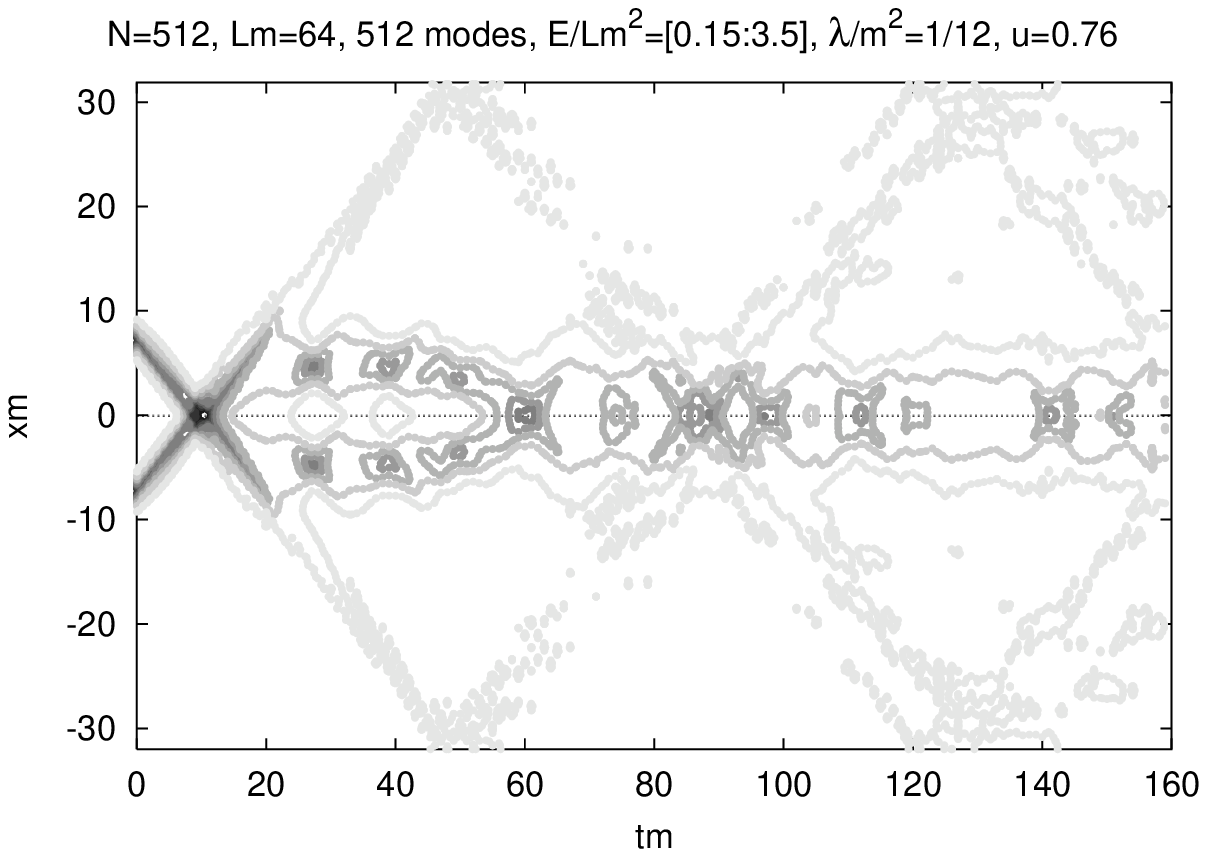}}
\subfigure[$u_i=0.765$ ($\gamma\approx1.553$).]{
    \includegraphics[width=0.475\textwidth]
    {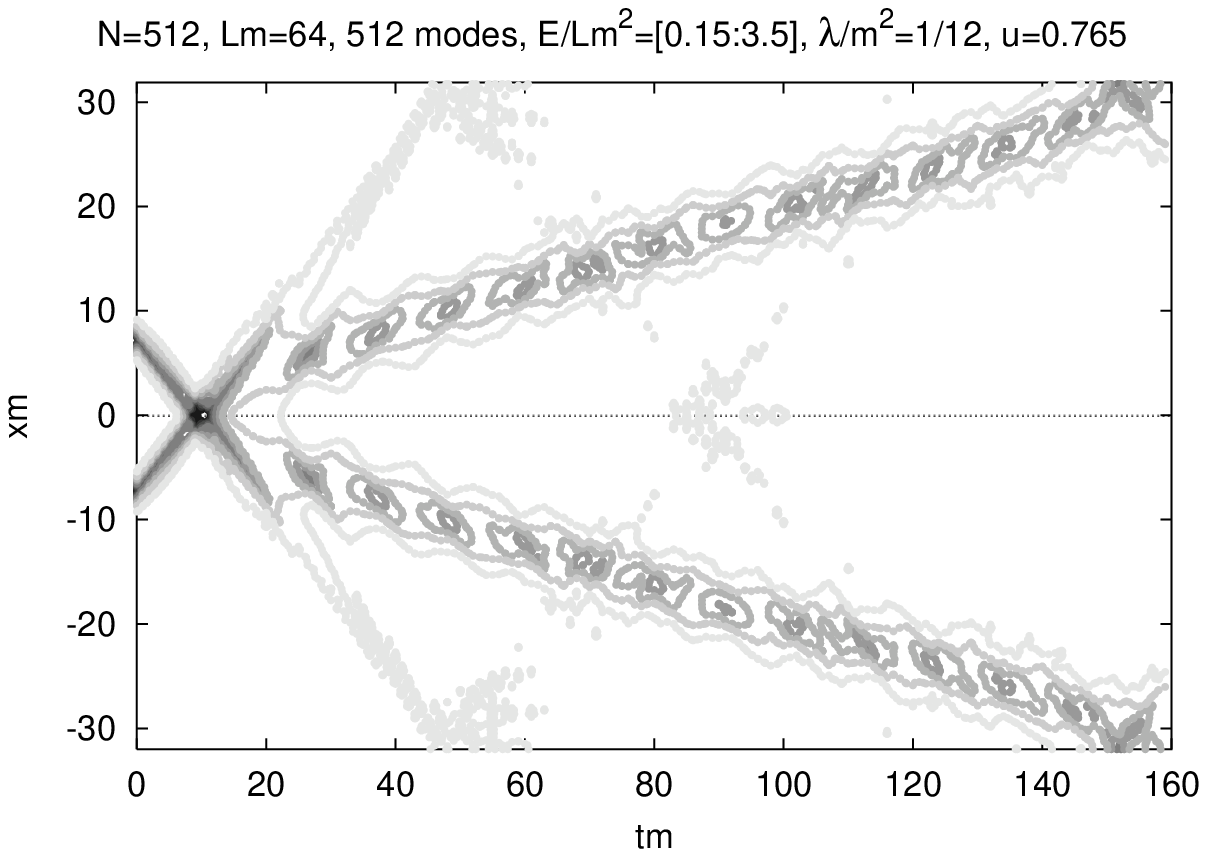}}
\subfigure[$u_i=0.770$ ($\gamma\approx1.567$).]{
    \includegraphics[width=0.475\textwidth]
    {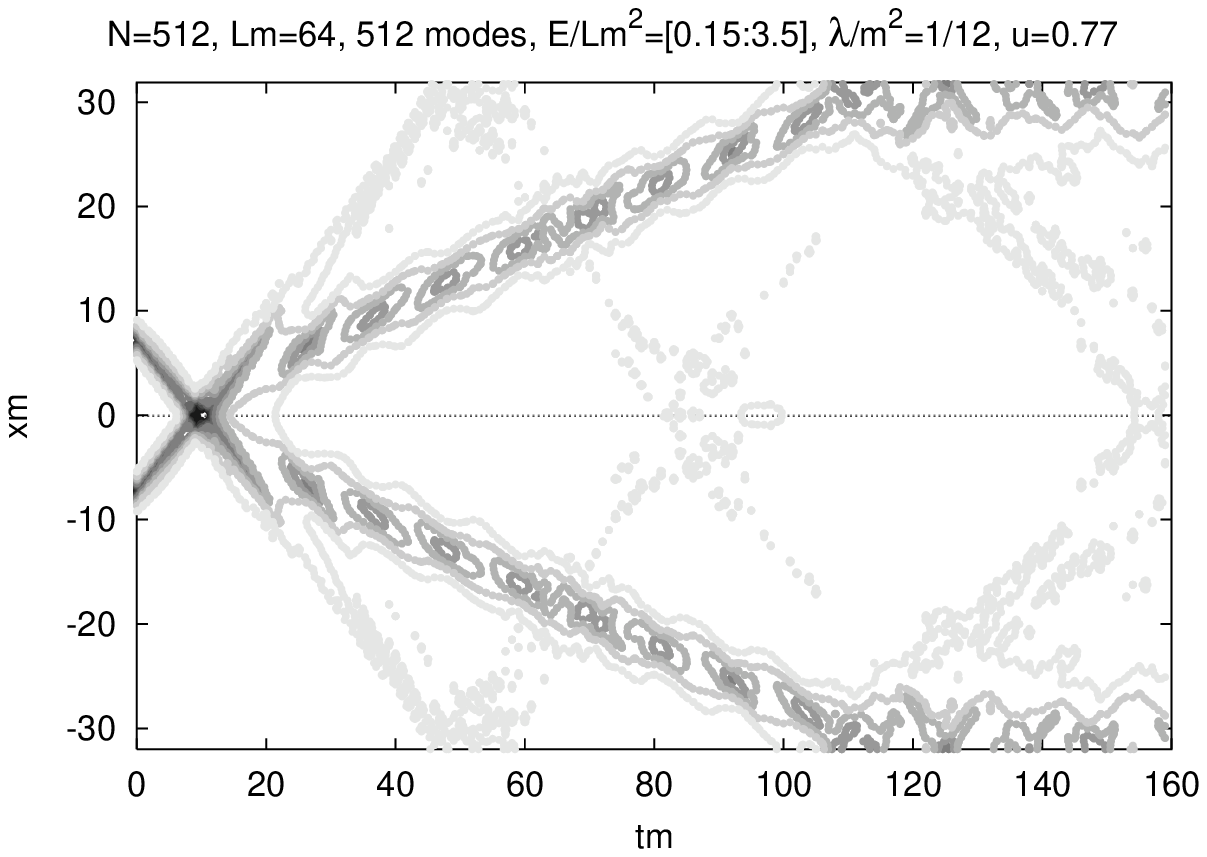}}
\subfigure[$u_i=0.800$ ($\gamma\approx1.667$).]{
    \includegraphics[width=0.475\textwidth]
    {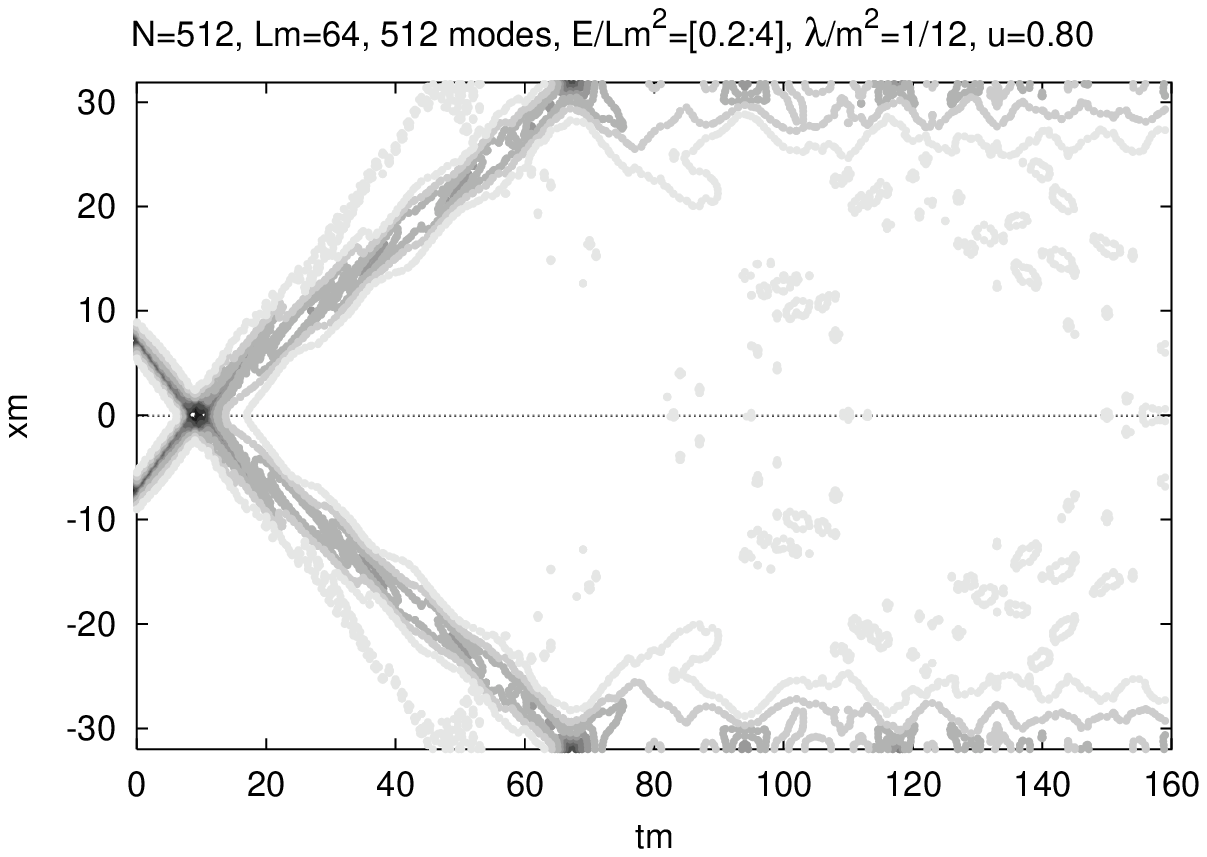}}
\end{center}
\caption{Colliding Hartree kink-antikink, $\lambda/m^2=1/12$, above and below
the critical speed.\label{fig:kinkcrithart}}
\end{figure*}
In Figs.~\ref{fig:kinkcrithart} the Hartree results for six different initial
speeds are shown. We see that a critical speed does indeed exist, with a value
somewhere between $0.760$ and $0.765$, indicating that the quantum kink pair is
less stable than its classical counterpart, at least at this coupling. One might
think that this instability is caused by the fact that we only have an
approximate quantum kink-antikink as initial condition, but as can be seen from
Figs.~\ref{fig:kinkcrithart}, the pair before the collision looks very stable,
there is hardly any wiggling, while after the collision the wave packet is
dispersing and oscillating and its speed has decreased considerably. The reason
for the higher instability is the radiative channel. This probably also causes
the fewer number of breather states found: only at an initial speed $u=0.760$,
very close to the critical speed, can we recognise an approximate breather
state. Bands of stability have not been found and in most simulations in which
the pair does not bounce back immediately, the radiation causes them to
annihilate.

\begin{figure}
\includegraphics[width=0.483\textwidth]{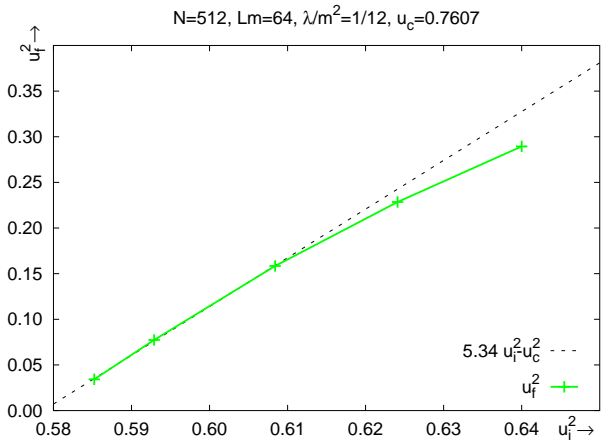}
\caption[$u_f^2$ as a function of $u_i^2$ for Hartree]{$u_f^2$ as a function of
$u_i^2$ in the Hartree approximation, showing roughly the same behaviour as in
classical theory, but with a $u_c=0.7607$.\label{fig:critspeed_hartree}}
\end{figure}

We also have investigated the functional behaviour of $u_f$ as a function of
$u_i$, to see if it has the same form \eqref{eq:critspeed_class}, valid in the
classical approximation. The result of several runs at different initial speeds
is shown in Fig.~\ref{fig:critspeed_hartree}. We see that close to the
critical speed, $u_f$ does indeed behave in the same way as in the classical
theory,
\begin{equation}
u_f^2 = 5.34 \left( u_i^2 - u_c^2 \right).
\label{eq:critspeed_hartree}
\end{equation}
The resulting critical speed $u_c$ can be determined very precisely in this way
and we obtain $u_c=0.760$. Further away from the critical speed the behaviour is
not linear in $u_i^2$, in contrast to the classical result, but the corrections
are relatively small.

We have done a few simulations using a stronger coupling $\lambda/m^2=1/6$ and
they indicate the general behaviour is only slightly different. For example,
the critical speed is not very different, it has a value between
$0.790$ and $0.795$, from a linear fit of $u_f^2$ versus $u_i^2$ we obtain
$u_c=0.793$ instead of
the value $0.7607$ at small coupling. Furthermore
at a
$u_i=0.80$ we find $u_f=0.280$, while at $\lambda/m^2=1/12$ we found
$u_f=0.538$.
Since $u_i=0.80$ is much closer to this value
of $u_c$, it explains why the value of $u_f$ at $u_i=0.80$ is a factor of 2
smaller. The prefactor in Eq.~\eqref{eq:critspeed_hartree} is higher at the
stronger coupling, around $7$ to $7.5$ instead of the $5.34$, but more data
would be needed in order to obtain an accurate answer.

Apart from the factor $3$ larger value of the critical speed, their are other
important differences between the classical and Hartree results,
Figs.~\ref{fig:kinkcrithart} and \ref{fig:kinkcritclass}. For example,
irrespectively if the pair annihilates or not, a lot of energy is radiated away
after the collision in the form of quasiparticles, moving with
approximately the speed of light. One might think that the radiation is mainly
described by the mode function contribution to the energy density, while the
energy density of the surviving kink pair comes from the mean field. However, by
checking the different contributions separately we found this is not the case,
the kinks are mainly described by the mean field, but also have a contribution
from the modes and the radiation is described by both together. Only the total
field is a physical quantity describing both the kink-antikink and the radiated
particles. Just as only the total two point functions describe the
quasiparticles which approximately thermalize.

\subsection{Numerical results: Scaling\label{ssec:scaling}}

\begin{figure*}[tbp]
\begin{center}
\subfigure[Hartree, $\lambda/m^2=1/1.25$, $\gamma=20$ ($u=0.9987$).]{
    \includegraphics[width=0.475\textwidth]{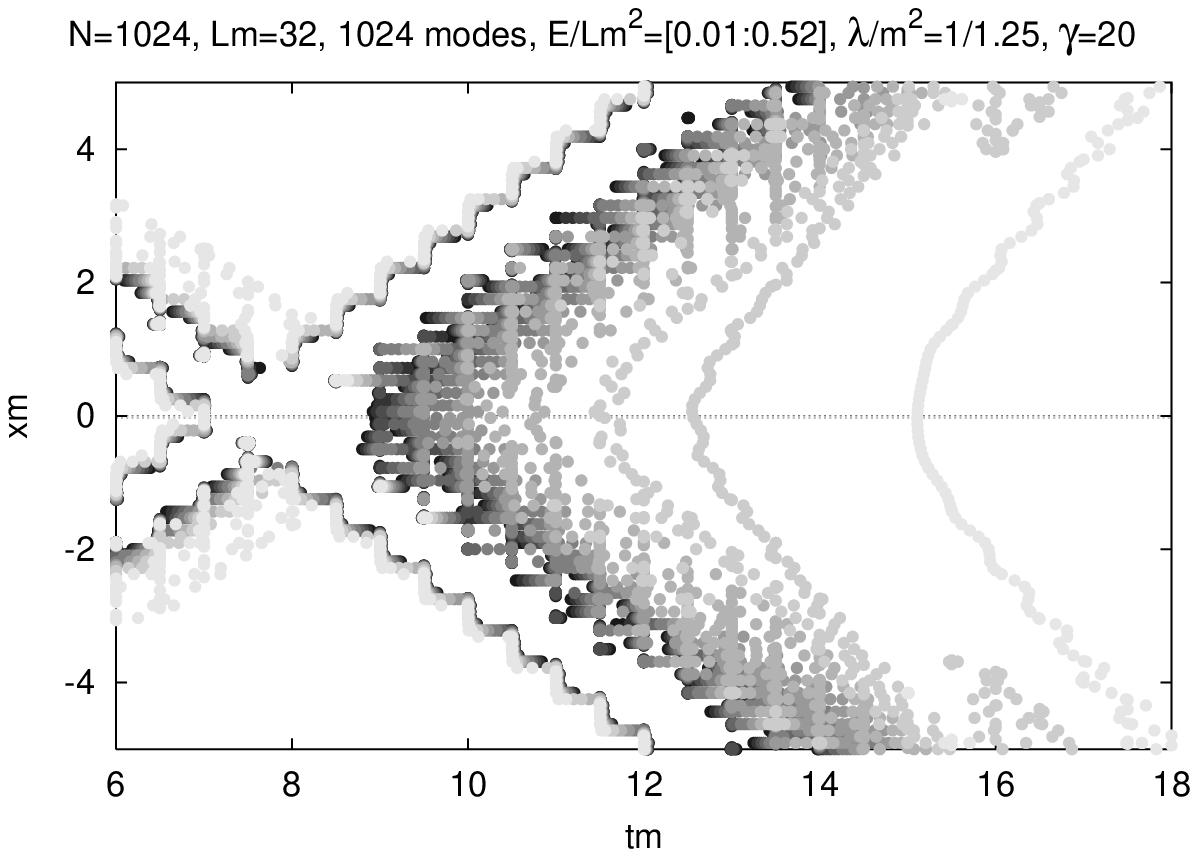}}
\subfigure[Classical, $\lambda/m^2=1/1.25$, $\gamma=20$ ($u=0.9987$).]{
    \includegraphics[width=0.475\textwidth]{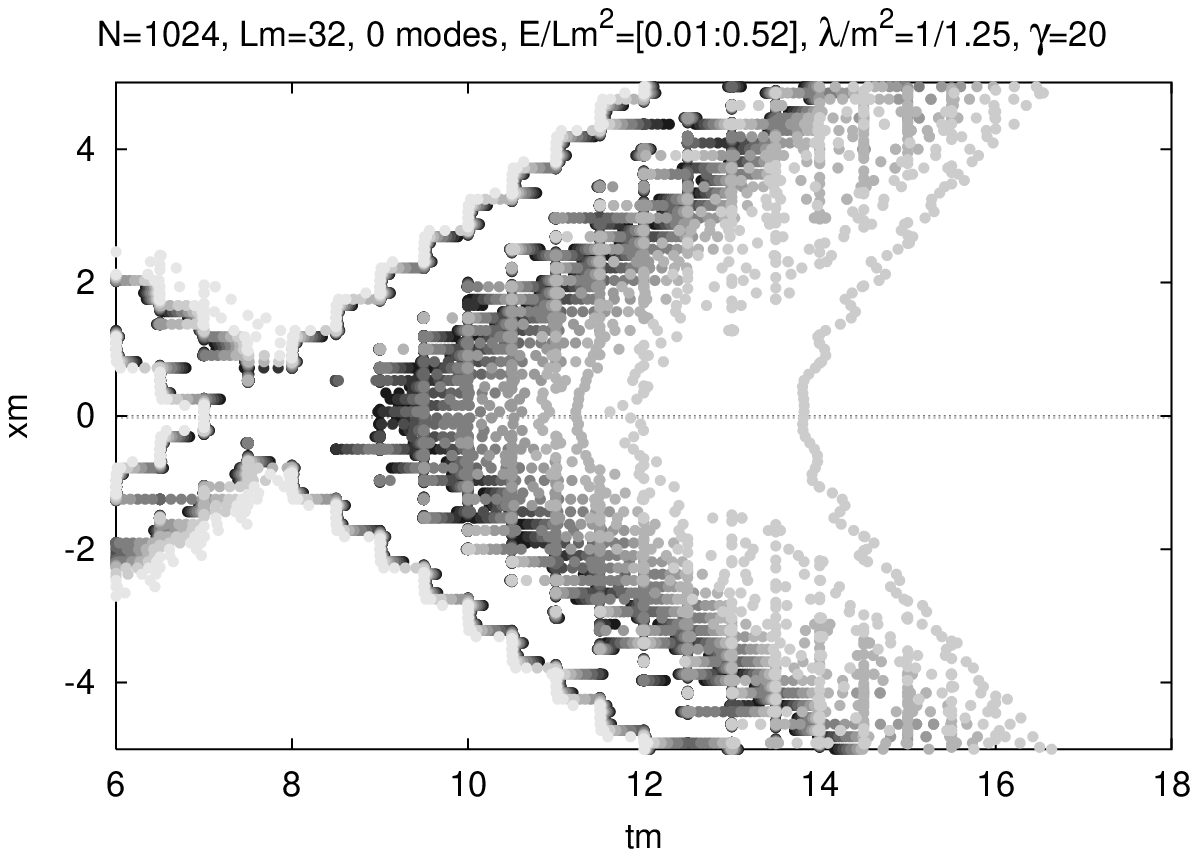}}
\subfigure[Hartree, $\lambda/m^2=1/1.25$, $\gamma=10$ ($u=0.9950$).]{
    \includegraphics[width=0.475\textwidth]{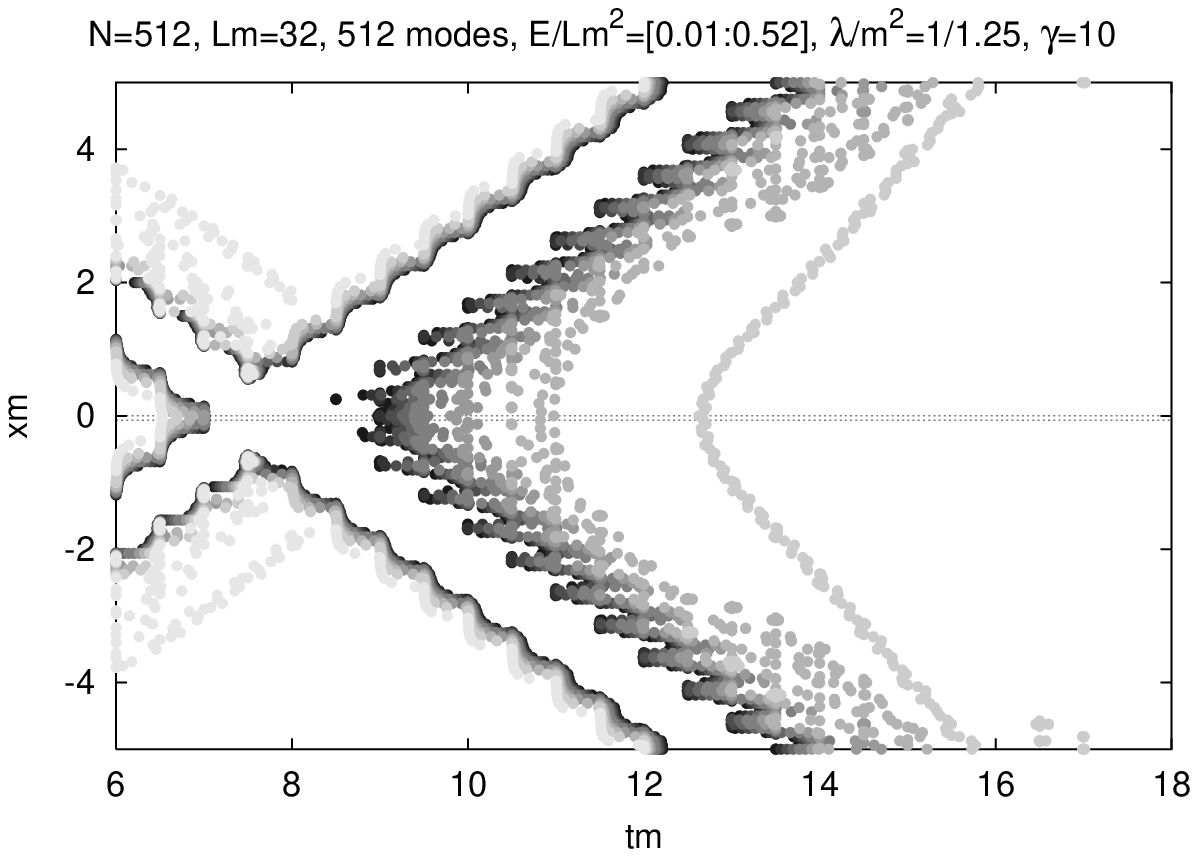}}
\subfigure[Hartree, $\lambda/m^2=1/12$, $\gamma=10$ ($u=0.9950$).]{
    \includegraphics[width=0.475\textwidth]{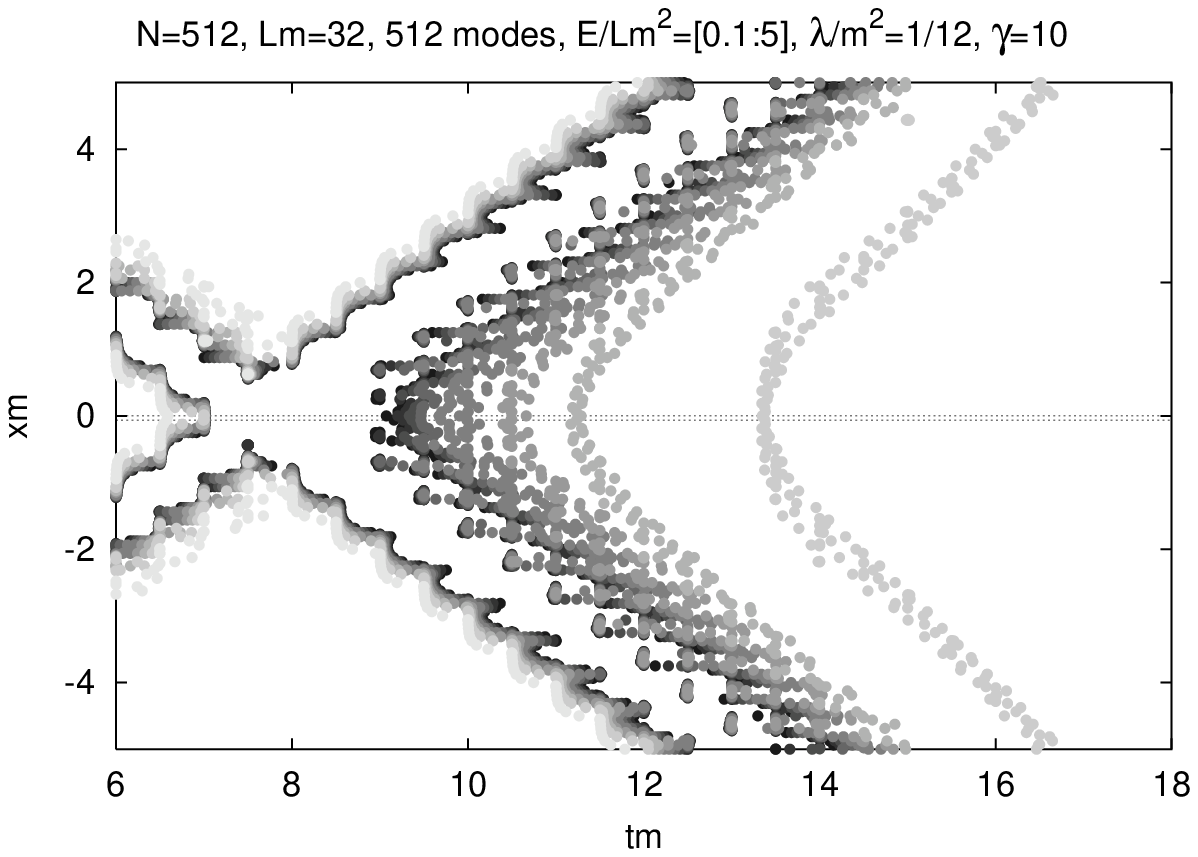}}
\end{center}
\caption{Colliding kink-antikink, period around the collision
only.\label{fig:plasmaregion}}
\end{figure*}

As we can describe kink-antikink collisions in a quantum theory, it is
interesting to study the possibility of using them as a toy model for heavy
ion collisions, such as has been conducted at the SPS, are currently being
conducted at RHIC and will later be at the LHC. We therefore look closer at the
region just after the actual impact, in collisions with high $\gamma$. In
Figs.~\ref{fig:plasmaregion} we show the results of four different simulations,
at different $\gamma$ factors and couplings, including one using classical
dynamics. 

\begin{figure}[tbp]
\subfigure[Hartree kink-antikink at $\gamma=20$ and $10$
($\lambda/m^2=1/1.25$).\label{fig:plasmaspeed}]{
    \includegraphics[width=0.44\textwidth]{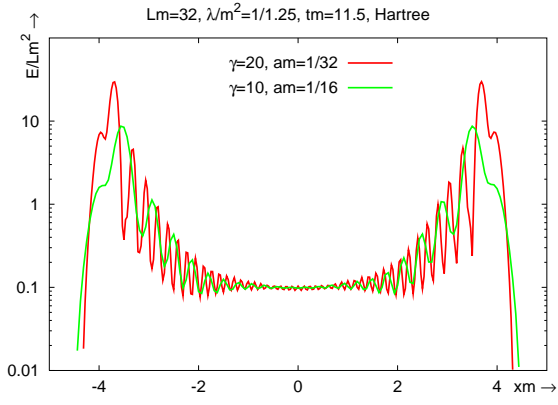}}
\subfigure[Hartree kink-antikink at $\lambda/m^2=1/12$ and $1/1.25$
($\gamma=10$). The energy density is normalised with the average energy density.
\label{fig:plasmacoupling}]{
    \includegraphics[width=0.44\textwidth]{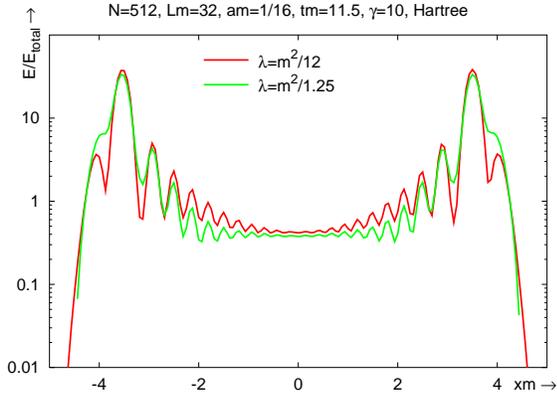}
}
\subfigure[Hartree mean field, Hartree total and classical
($\lambda/m^2=1/1.25$, $\gamma=20$). \label{fig:plasmaclashart}]{
    \includegraphics[width=0.43\textwidth]{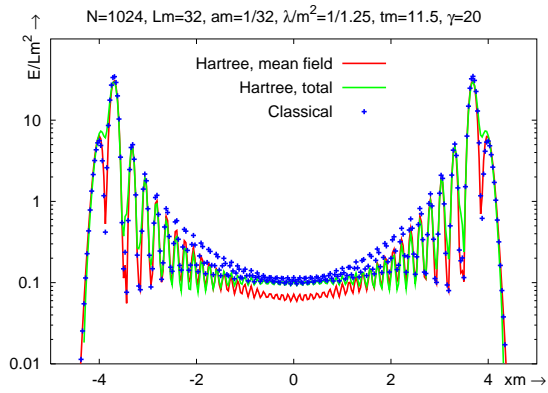}
}
\caption{Comparison of energy density profiles at a time-slice $tm=11.5$,
(collision at $tm=8$) for different parameters.\label{fig:timeslice}}
\end{figure}

The plots show a remarkable similarity. In order to investigate this similarity
more quantitatively we made time slices shortly after the impact, shown in
Figs.~\ref{fig:timeslice}, at a time $11.5$, the impact itself was
at a
time $tm
\approx 8$ (the determination of the exact time of impact has some complications
to which we will return later).

In Fig.~\ref{fig:plasmaspeed} we see that a different impact velocity only
influences the resulting energy density in the emerging kink-antikink, the
central region is the same for both speeds. There, the difference in initial
speed only shows up in the oscillation wavelength, which scales due to the
$\gamma$ factor.

\begin{figure}[tbp]
\includegraphics[width=0.483\textwidth]{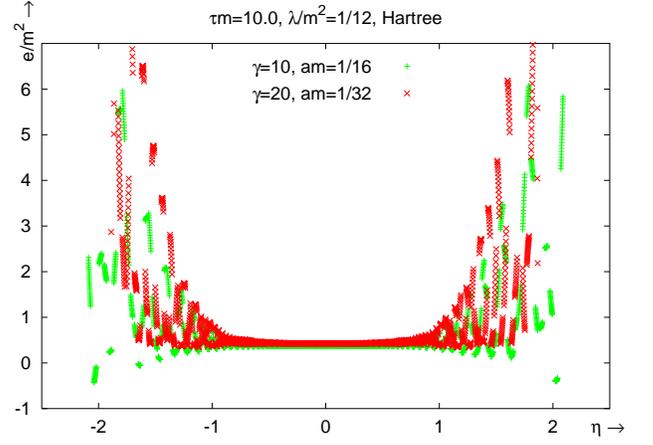}
\caption{Total energy density in ``Bjorken variables'' $\tau$ and $\eta$ for two
different initial speeds. Note the linear scale.\label{fig:tau_eta}}
\end{figure}

In Fig.~\ref{fig:tau_eta} we have plotted a similar comparison at smaller
coupling, $\lambda/m^2=1/12$, as a function of 
space-time
rapidity $\eta=\arctanh(z/t)$ at
constant ``proper time'' $\tau=\sqrt{t^2-x^2}=10 m^{-1}$ ($t=0$ at the
collision). Even on the linear vertical scale, the plateau heights are almost
identical. It is also striking to see how flat the plateau is, an indication for
hydrodynamic scaling. Note that $xm \approx 10 \, \sinh(1) \approx 12$ for the
edges of the plateau, a large distance compared to the size of the kink, which
is much smaller than $1$.

In Fig.~\ref{fig:plasmacoupling} we show the difference between the two
couplings. We have rescaled the energy density with its average over the whole
system, in order to facilitate the comparison. Again, the difference is small.
At the stronger coupling the energy is slightly more concentrated in the
kink-antikink.

Finally in Fig.~\ref{fig:plasmaclashart} we compare the difference between
Hartree and classical dynamics. In both simulations the kink-antikink pair
region is very similar. The central region shows some differences, although the
total Hartree energy density in this region is very close to the classical
energy density. The modes are essential for this, the mean field is more
concentrated in the kink-antikink.

To summarize, we have shown that the collisions show scaling behaviour for an
extensive range of energies and couplings: they result in approximately the same
central region, which is very flat in
space-time
rapidity. Furthermore at high incident
speeds the difference between Hartree and classical is relatively small:
classical dynamics can be used in the study of these collisions, giving
foundation for its use in the study of heavy ion collisions. 

It is interesting to compare the results obtained here in kink-antikink
collisions with what is known about heavy ion collisions. It is already long
accepted \cite{Bj83} that the height of the central plateau only weakly depends
on the incident speed. Experiments at RHIC, at center of mass energies of $130$
and $200 \, GeV$ per nucleus (i.e., $\gamma=130$ and $\gamma=200$,
respectively), indicate an increase in particle density of $14 \pm 5\%$
\cite{Ba01}. At the smaller coupling of $\lambda/m^2=1/12$ we find, at a time
$tm=3$ after the collision, an increase of only about $3\%$ when increasing from
$\gamma=10$ to $\gamma=20$. However, given the very different physical systems,
a better match would be a mere coincidence. Longer after the collision, $m
\tau=10,\dots,100$, the difference in energy density between the two simulations
becomes somewhat larger, as we will see later.

Furthermore, our findings, that only a relatively small fraction of the energy
remain in the central region and the concept of a central plateau in
space-time
rapidity
are, at least qualitatively, consistent with real-life heavy ion collisions. A
quantitative comparison would not be very meaningful.

\begin{figure*}[tbp]
\begin{center}
\subfigure[Log-log plot of energy density at $x=0$.\label{fig:plasmalineclass}]{
    \includegraphics[width=0.475\textwidth]{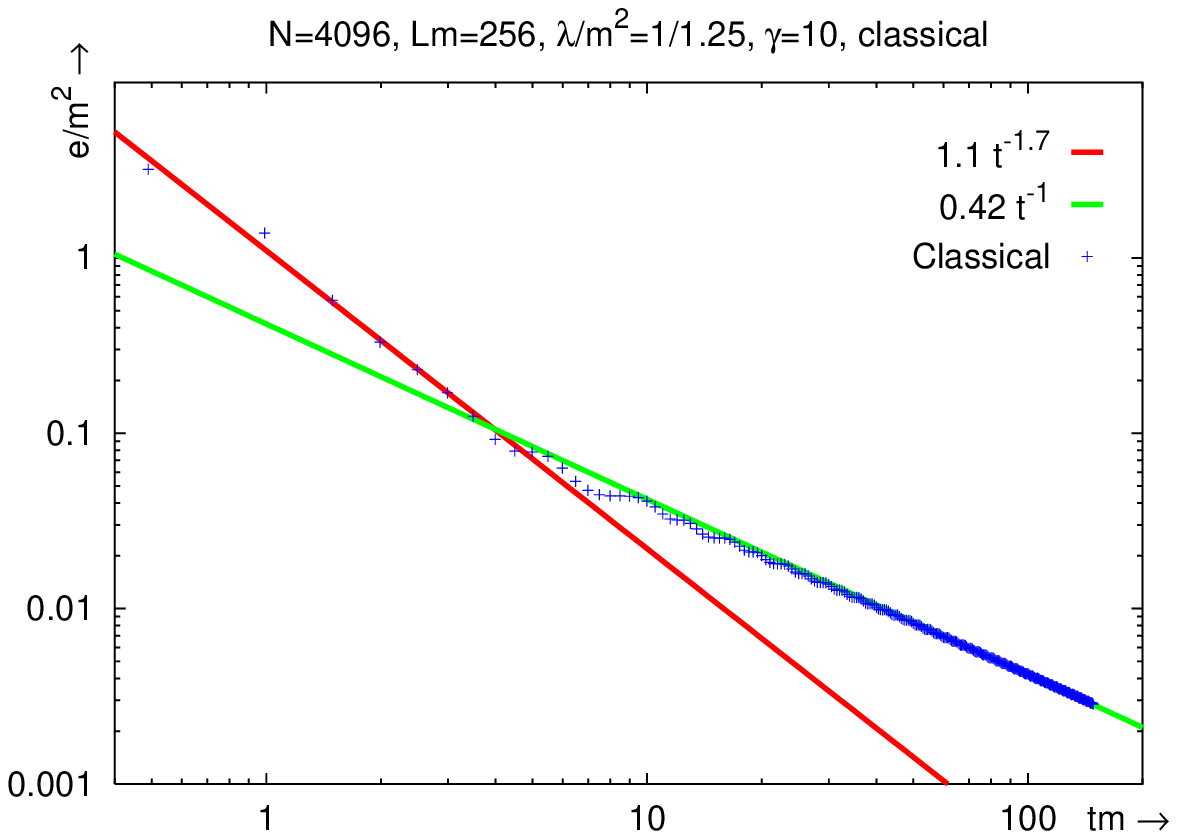}}
\subfigure[Pressure $p=e-2V$ and energy density. Note that $e
$ falls right on top of $p$.\label{fig:pressureclass}]{
    \includegraphics[width=0.475\textwidth]{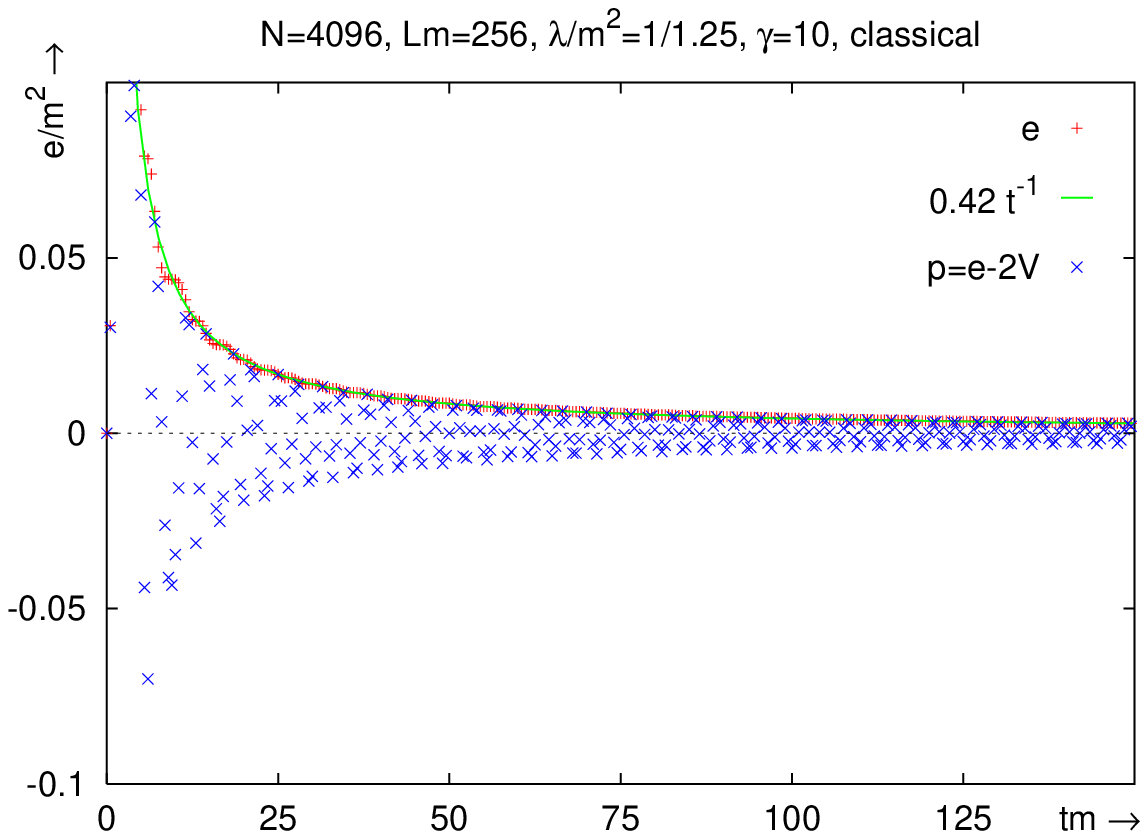}}
\end{center}
\caption[Energy density in the central region]{Energy density and
pressure in the central region, after a classical kink-antikink collision. At
later times, both are consistent with $p=\text{constant}$ and $c_0=0$.}
\end{figure*}
A first study of hydrodynamic scaling in a $\phi^4$ theory can be found in
Ref.~\cite{BeCo01}. The authors calculate the energy momentum tensor, which
in our metric equals
\begin{equation}
T_{\mu \nu} = \partial_\mu \phi \partial_\nu \phi + \eta_{\mu \nu} \mathcal{L},
\end{equation}
in two systems, a colliding kink-antikink\footnote{Note that in
Ref.~\cite{BeCo01} a product of a kink and antikink is taken, while we use a sum
of the two, Eq.~\eqref{eq:iniclass}. Of course both are approximate
kink-antikink solutions and the difference should be small.} and a decaying
Gaussian wave packet. For a perfect fluid the energy momentum tensor can be
expressed in the energy density and pressure. The assumption of a perfect fluid
is valid when collisions can be neglected, as in the (homogeneous) Hartree
approximation. For example, the trace of $T_{\mu\nu}$ gives (in $1+1$ dimensions)
\begin{subequations}
\begin{align}
T_\mu^\mu &= -e+p = -2 V, \\
e&=\frac{1}{2} (\partial_t \phi)^2 + \frac{1}{2} (\partial_x \phi)^2 + V, \\
V&=\frac{1}{2}\mu^2 \phi^2 + \frac{1}{4} \lambda \phi^4 + \frac{\mu^2}{\lambda}.
\end{align}
\label{eq:traceTmunu}
\end{subequations}
Note that this is a slightly different definition from the one given in
Ref.~\cite{BeCo01}, Eq.~(13), as we would like both $T_{00}$ and $T_{11}$ to
vanish in the vacuum $\phi=v$.
Also note that we just write the classical expressions, the extension to Hartree
is straightforward.
From Eqs.~\eqref{eq:traceTmunu} we find the
pressure
\begin{equation}
p=\frac{1}{2} (\partial_t \phi)^2 + \frac{1}{2} (\partial_x \phi)^2 - V.
\end{equation}
Using the scaling behaviour one can then, for example, express the speed of sound
in the energy density in the center
\begin{equation}
e \propto \tau^{-(1+c_0^2)} \to t^{-(1+c_0^2)} \qquad \text{at $x=0$,}
\label{eq:soundspeed}
\end{equation}
\begin{figure}[tbp]
\includegraphics[width=0.483\textwidth]{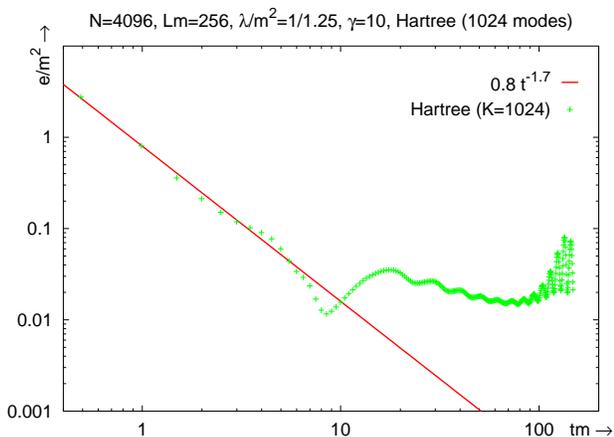}
\caption{As in Fig.~\ref{fig:plasmalineclass} but now for a Hartree
kink-antikink collision.\label{fig:plasmalinehartree}}
\end{figure}
\begin{figure*}[tbp]
\begin{center}
\subfigure[Classical.\label{fig:plasmaline12class}]{
    \includegraphics[width=0.47\textwidth]{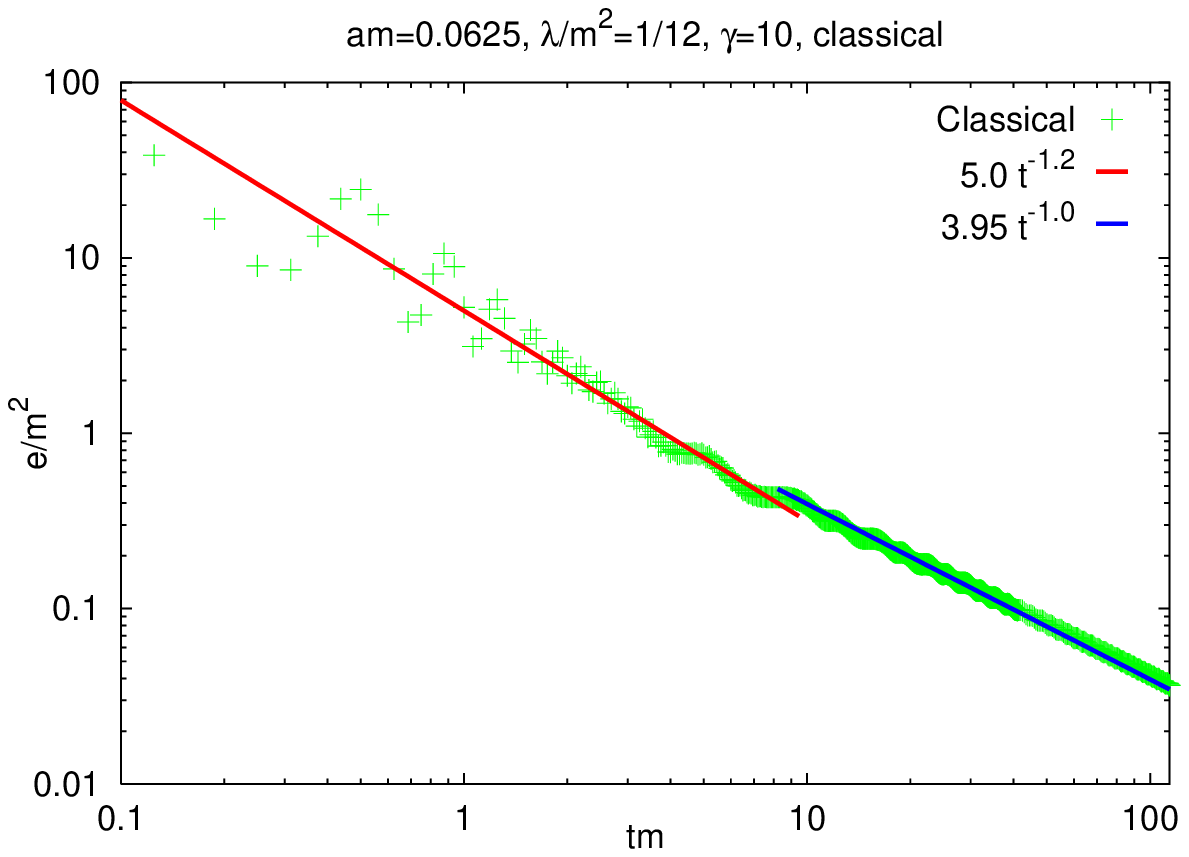}}
\subfigure[Hartree.\label{fig:plasmaline12Hartree}]{
    \includegraphics[width=0.47\textwidth]{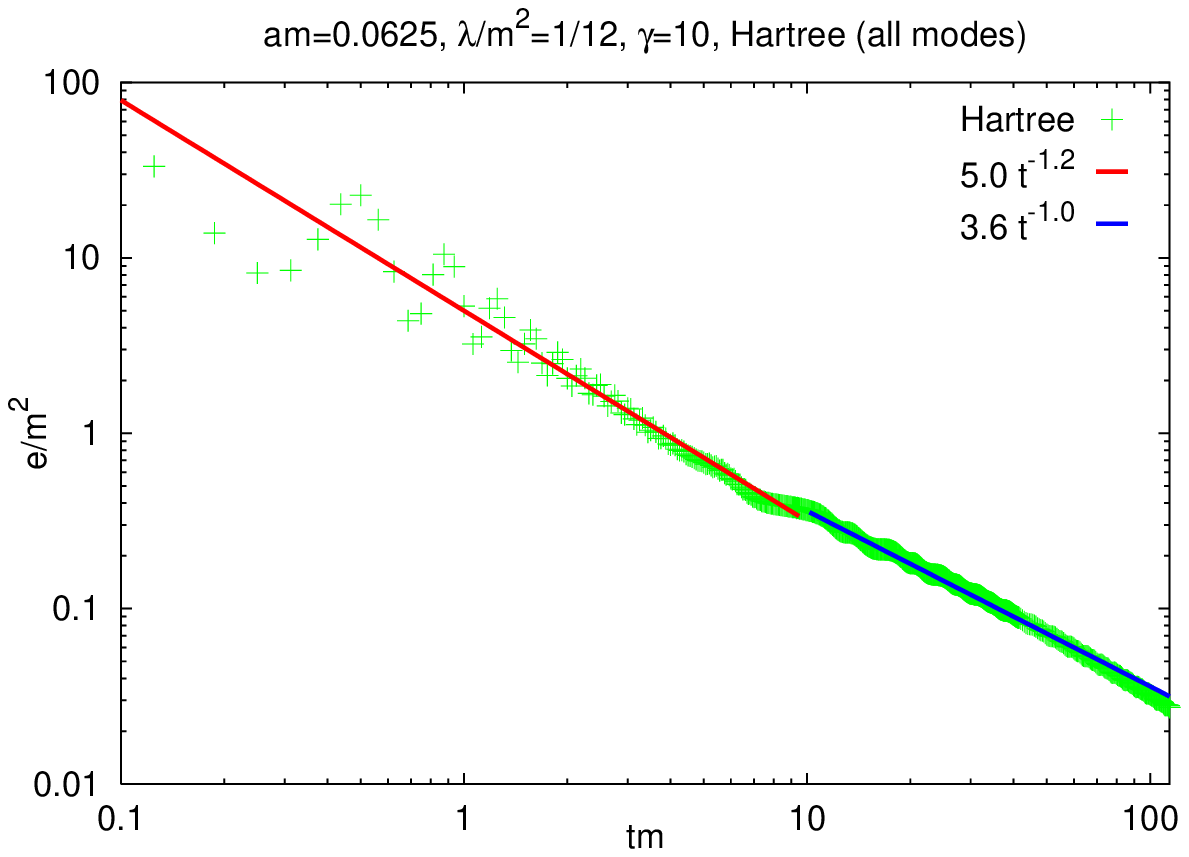}}
\end{center}
\caption{As in Fig.~\ref{fig:plasmalineclass} and \ref{fig:plasmalinehartree}
but now for $\lambda/m^2=1/12$.\label{fig:plasmaline12}}
\end{figure*}
where $\tau=\sqrt{t^2-x^2}$ is the ``proper time'' and $c_0$ is the speed of
sound. For a classical kink-antikink collision at strong coupling
$\lambda/m^2=1/1.25$ we plotted the central energy density as a function of time
(where the time is set to zero at the \emph{end} of the collision), on a log-log
plot in Fig.~\ref{fig:plasmalineclass} and, on a linear scale, together with the
corresponding pressure, in Fig.~\ref{fig:pressureclass}. Both from
Eqs.~\eqref{eq:traceTmunu} and Eqs.~\eqref{eq:soundspeed} we find that shortly
after the collision, the pressure in the central region becomes zero, leading to
a vanishing speed of sound. This indicates that after this initial stage, the
interaction becomes negligible and we just see free expansion, in agreement
with the flat central plateau. Initially, however, the system is interacting and
there is a non-vanishing speed of sound of about $0.8$. The actual value depends
quite strongly on the time origin. We have derived the value to use, from
contour plots such as Figs.~\ref{fig:plasmaregion}. They show that the
kink-antikink stay together for a short period of time $\Delta t m\approx 0.5$.

The result obtained from Hartree dynamics is shown in
Fig.~\ref{fig:plasmalinehartree}. In this case we can only find a speed of
sound in the first stage, resulting in a
value
very similar to the
classical result. The second stage exhibits strange oscillations but no power
behaviour. This is caused by the emergence of a ``symmetric minimum'' at
$\phi=0$, already discussed in Sec.~\ref{ssec:kinkdecay}. In
Figs.~\ref{fig:plasmaline12} we show the results for collisions at the smaller
coupling $\lambda/m^2=1/12$. Here we see that both classical and Hartree
results give an approximately vanishing speed of sound and thus pressure. The
initial speed of sound is for both smaller, about a factor of
$2$ compared to the strong coupling,
which is
(qualitatively) consistent with a smaller coupling. The Hartree result,
Fig.~\ref{fig:plasmaline12Hartree}, shows small deviations at the latest times,
indicating some interaction. At a higher incident speed, $\gamma=20$, the
prefactor in the $1/\tau$ decay rises to $4.4$, both for the classical and
Hartree simulations. It is interesting to see that this increase is actually
very close to the RHIC results as discussed above.

It is difficult to compare our results quantitatively with those of Bettencourt
et al.~\cite{BeCo01}, since their result for the speed of sound was obtained for
a disintegrating Gaussian wave packet in the ``symmetric phase.'' The effective
coupling in the ``symmetric phase'' is much smaller than in the ``broken phase''
\cite{SaSm02} and we have studied only the central region in the wake of a
kink-antikink collision, as this is more similar to a heavy ion collision.
The free expanding region is not present in the results of Ref.~\cite{BeCo01}
and the power behaviour is less pronounced. However, the speed of sound in the
initial region is very comparable.

\section{Thermal kink nucleation\label{sec:kink_nucl}}

In this section we briefly discuss the thermal creation and annihilation
properties of the kink-antikink pairs. There is a long history of papers on the
subject of thermal nucleation. See, for example, Ref.~\cite{CuKr80} and
references therein for an analytical study, and for example
Refs.~\cite{GrRu88,AlFe92,AlHa93a,AlHa93b,HaLy00} for numerical studies.
Very recently Ref.~\cite{Ho03} has
made a study
in a heavy ion collision
context, calculating classically the multiplicities of $K\bar{K}$ in an
expanding Bjorken frame, starting from a Boltzmann distribution.

All these (numerical) studies so far have only considered the classical
nucleation rate. The Hartree ensemble approximation method allows us to study
the creation of kink-antikink pairs starting from a thermal Bose-Einstein
distribution and to compare the classical and Hartree
approximation. One of the problems we are thus faced with is a proper
definition of kink number. Since only pairs, with no net kink number can be
created, we need an effective kink number definition. The actual winding number
will always be $0$ or $1$, depending only on the boundary conditions.
Furthermore, in the quantum theory, one might think the kinks will be fully
described by the mean field, while the mode functions just describe fluctuations
around them. However, as we already noticed before, this split-up cannot be
taken so rigorously: the kink number will also be partially described by the
modes
and
a definition based on the ensemble description of our system is needed,
as only ensemble averaged quantities are physically meaningful.

\begin{figure}[tbp]
\begin{center}
\includegraphics[width=0.483\textwidth]{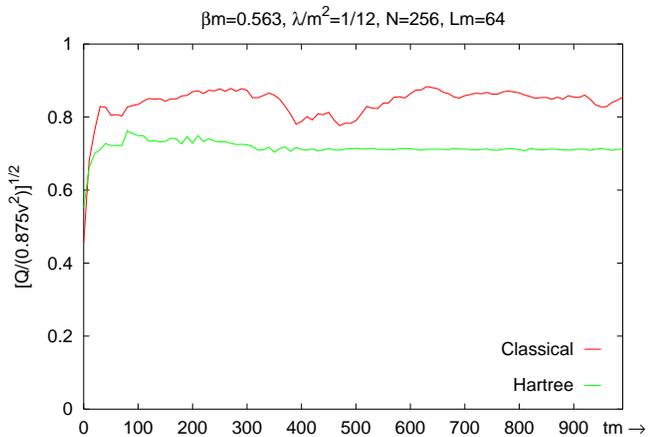}
\end{center}
\caption[Kink indicator $Q$ for classical and Hartree dynamics]{Kink indicator 
parameter $Q$ from Eq.~\eqref{eq:kinkorderparam} for classical dynamics
(upper plot) and Hartree dynamics (lower plot). Both values are normalised with
the value $0.875$, the value of $Q$ for a symmetrically placed kink-antikink
pair in a volume $Lm=64$. Of course this value heavily depends on the exact
position of the kinks.\label{fig:kinkorderparam}}
\end{figure}

\begin{figure*}[tbp]
\subfigure[$tm=1$]{\includegraphics[width=0.46\textwidth]{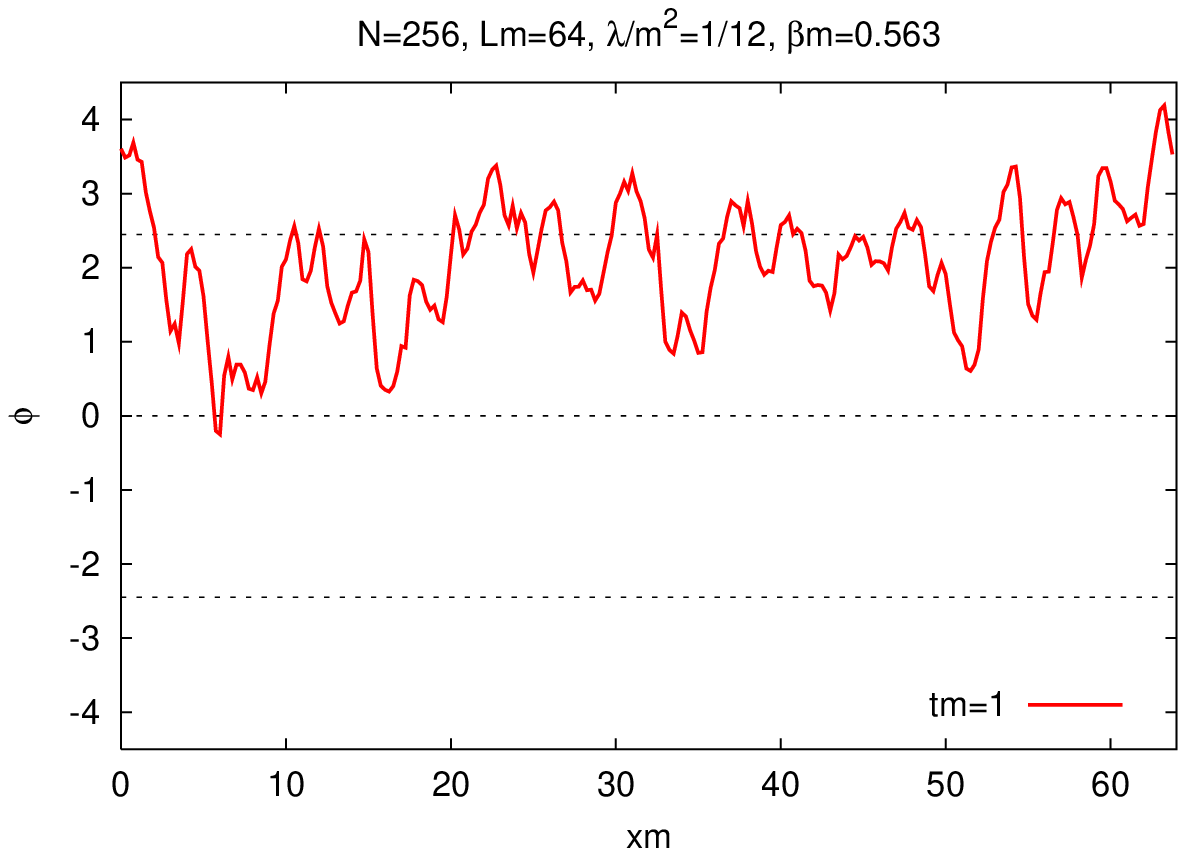}}
\subfigure[$tm=100$]{\includegraphics[width=0.46\textwidth]{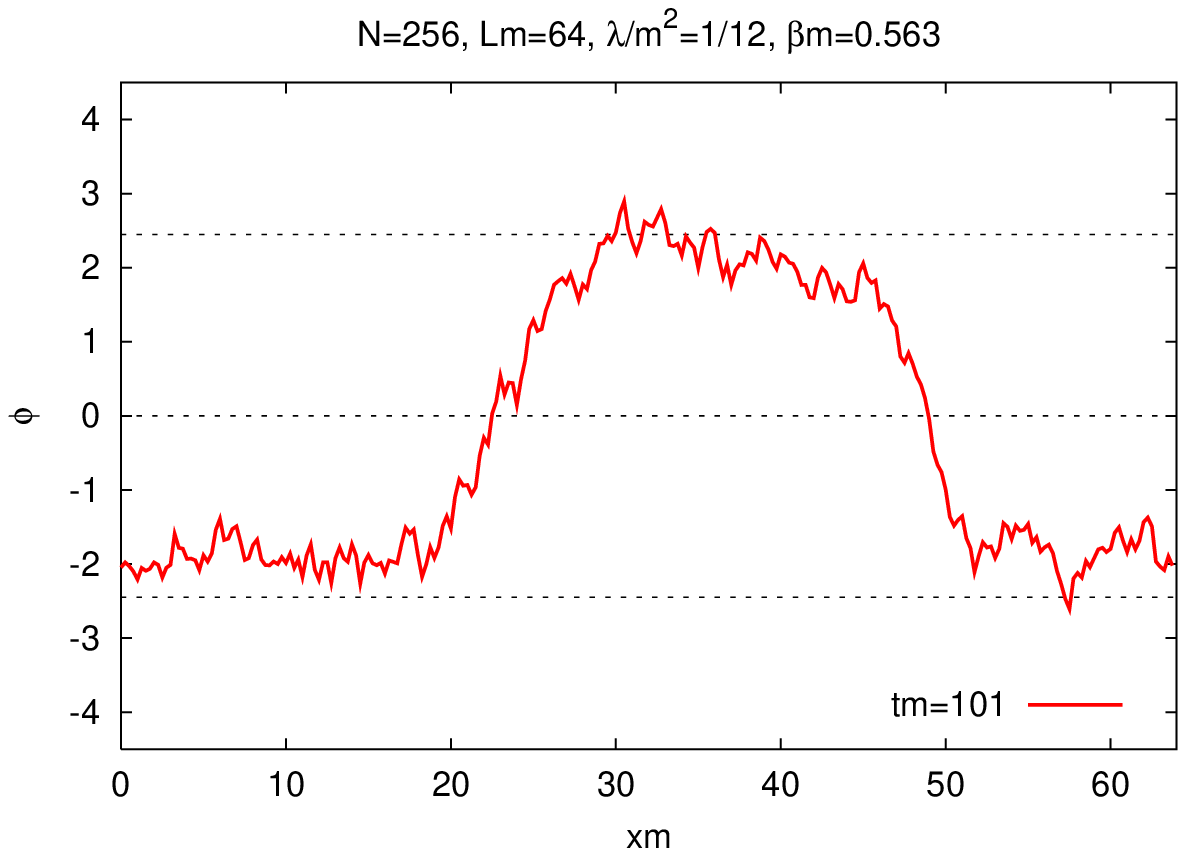}}
\subfigure[$tm=160$]{\includegraphics[width=0.46\textwidth]{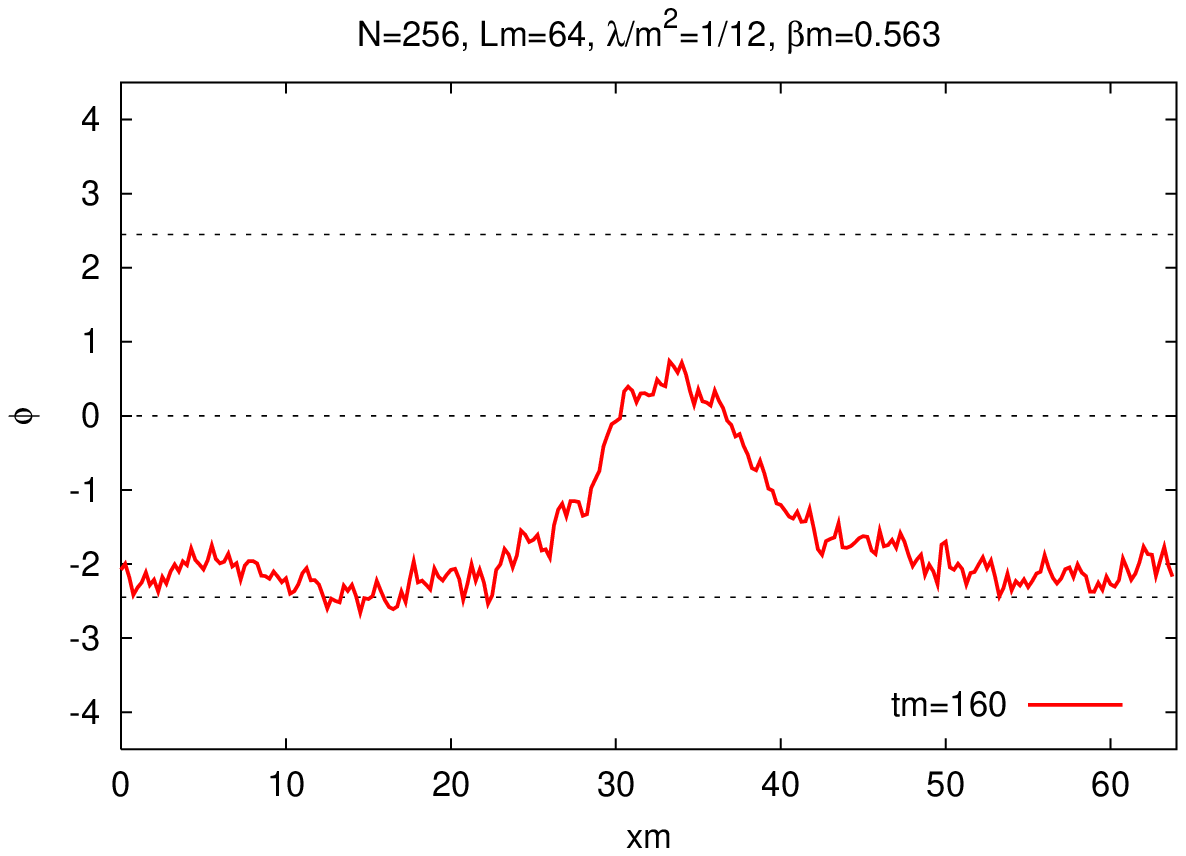}}
\subfigure[$tm=165$]{\includegraphics[width=0.46\textwidth]{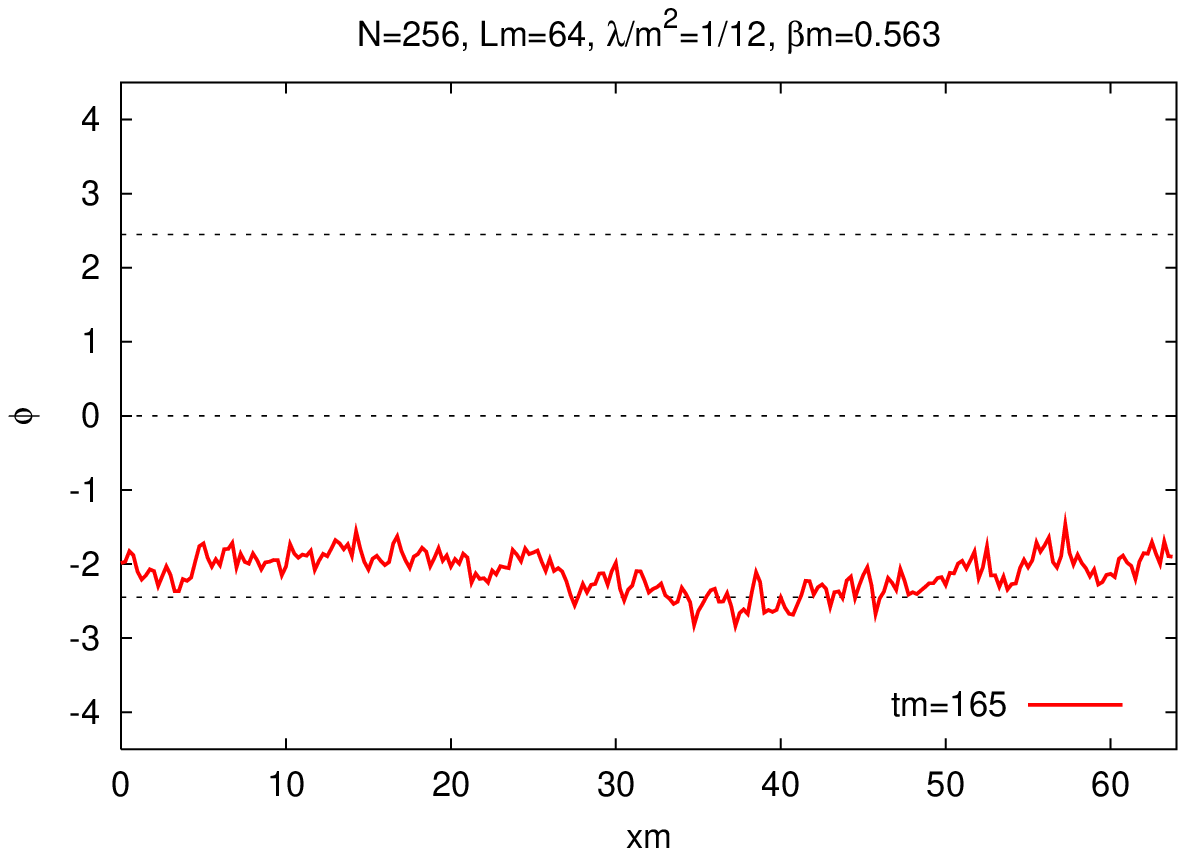}}
\caption[Thermal Hartree kink nucleation and annihilation]{Thermal kink
nucleation and subsequent annihilation in the Hartree approximation. The dotted
lines are at the zero temperature ``broken'' and ``symmetric
minima'' $\pm v(T=0)$ and $0$.\label{fig:kinkexample}}
\end{figure*}

In the classical theory, a useful quantity to look at is the following,
\begin{equation}
Q_{\text{class}} = \overline{\phi(x)^2} - \overline{\phi(x)}^2,
\end{equation}
where the overline denotes a spatial average over the full lattice size. When a
kink-antikink pair is present, this observable will become of the order $v^2$.
If the coupling is not too small, it will therefore give a reasonable indication
for their presence. Note that it cannot distinguish between one or multiple
pairs and that it becomes smaller if the pair is not at maximum separation. The
advantage is that we can easily extend its definition to the quantum theory
\begin{equation}
Q =
\overline{\braket{\hat{\phi}(x) \hat{\phi}(x)}} -
\overline{\braket{\hat{\phi}(x)}}\;\overline{\braket{\hat{\phi}(x)}}
\end{equation}
and that it can be easily found from $S_k$, Eq.~\eqref{eq:Sk}. It is however UV
divergent and should be renormalized by subtracting the vacuum contribution:
\begin{equation}
Q_{\text{ren}} = \frac{1}{L} \sum_k
\Bigl( S_k - \frac{1}{2 \omega_k^{\text{free}}} \Bigr).
\label{eq:kinkorderparam}
\end{equation}
We now have a quantum kink indicator, showing if a pair is present and which
can be easily compared to the classical theory. This is done in two
simulations at $\lambda/m^2=1/12$ in a volume $Lm=64$, at an inverse temperature
$\beta m=0.563$, just below the thermal phase transition at $\beta m=0.562$. At
such a high temperature we expect the highest creation rate. In a volume
$Lm=64$, it would follow from Refs.~\cite{AlHa93a,AlHa93b}, that we should find
about a half to two kinks per total volume, depending on the counting algorithm.
The result for $Q$ is shown in Fig.~\ref{fig:kinkorderparam}. 
We can clearly see it is different from zero, showing the presence of
kink-antikink pairs. Furthermore, these pairs emerge through the dynamics,
initially the number is lower. Finally, although the quantum number is smaller
than the classical, it does \emph{not} go away, while, if we look at the mean
fields in different realisations separately, we find the kink-antikinks do seem
to disappear in the Hartree approximation. From the $Q$ indicator we conclude
that they first are contained in the mean field, while later their description
is taken over by the modes, again showing that only the total two-point
functions, from the total field, describe the physical quantum field. As an
example of their emergence from the dynamics, we plotted, in
Fig.~\ref{fig:kinkexample}, the mean field of one of the realisations at 4
different times. Initially the field fluctuates around one of its minima, with a
reasonable amount of energy, it then forms a $K\overline{K}$ pair, which
subsequently annihilates, while transferring its energy to the quantum modes, as
can be seen from the smaller amplitude of the fluctuations. The total
two-point
function still describes $K\overline{K}$ pairs, but this cannot be seen from the
separate realisations.

\section{Conclusion\label{sec:conclus}}

In this paper we looked at the topological defects which can be present in a
scalar $\lambda \phi^4$ theory: kinks and antikinks. After rederiving the
classical solutions, we used both classical and Hartree dynamics in studying
their annihilation, both when initially at rest and when boosted. We found that
classical kink pairs are much more stable than their Hartree counterparts.
Hartree kink pairs at smaller coupling are therefore also more stable than at
larger couplings. Although an exact Hartree solution to the equations of motion
is not known, by damping the mean field equation, we
are able to
find an approximate
solution. This damping can, especially at larger couplings, prolong the survival
time enormously, but is still orders of magnitude shorter than for a classical
kink-antikink pair, due to radiation in the form of quantum particles, the
quasiparticles discussed in Refs.~\cite{SaSm00a,SaSm00b}. We suggested an
algorithm to find an accurate, numerical and more stable solution for the
Hartree equations of motion, but have not yet implemented it.

Using a damped mean field equation of motion we were also able to determine
approximately the quantum kink mass. The Hartree approximation seems to give
better results than the one-loop semiclassical approximation, especially at
larger couplings, where the semiclassical approximation becomes less reliable.
The mean field contribution shows a funny increase as a function of coupling,
when approaching the phase transition, although the total kink mass behaves
regular. It is not yet clear how to interpret this. Damping the equations is not
very efficient, as it can only be implemented in the mean field equation,
leading to very long relaxation times. We found a rapid short relaxation until a
time $tm\approx10-20$, in which most of the mean field energy is removed,
followed by an exponentially slow relaxation in which the energy of the modes is
drained indirectly via the mean field. Comparing the Hartree results with Monte
Carlo results from Refs.~\cite{CiTa94,ArWi99} seems to indicate that the
Hartree results, although larger than the one-loop semiclassical results, are
still too small.

We have shown that a colliding kink-antikink pair, after annihilation, also
leads to an approximate thermal Bose-Einstein spectrum, just as a flat initial
ensemble. The initial energy distribution has a form which approaches the
thermal distribution more easily than the flat ensemble, as used in
Ref.~\cite{SaSm00a}. This makes it possible to recognise the Bose-Einstein
distribution already in an early stage. However the statistical errors are
larger, since we cannot average over multiple initial conditions. It does show
that the result of Ref.~\cite{SaSm00a} is rigorous, it does not depend on the
specifics of the initial ensemble.

In the study of kink collisions, we found once more that the classical kinks are
much more stable: for the classical dynamics, we reproduced the results of
Ref.~\cite{CaSc83} on the critical speed and the existence of approximate
breather modes. In the Hartree approximation the critical speed is considerably
higher and we were not able to find stability bands, i.e., approximate breather
solutions. In order to make this result rigorous, many more simulations have to
be done, at higher numerical precision, since the radiation makes it difficult
to see clearly if a short-lived bound state, or breather, has emerged. The
precise dependence on the coupling is also still an open question. In the
classical theory the dimensionless critical speed is independent of $\lambda$ as
the action can be rewritten in a way that $\mu=\lambda=1$, while in the quantum
theory, this is not possible.

We compared some of the results from these kink-antikink collisions with the
heavy ion collisions such as are currently carried out at the RHIC in Brookhaven
and in the future at the LHC at CERN in Gen\`eve. Although the scalar $\phi^4$
theory is too simple to serve as a toy model for this, it is interesting to see
that we can certainly compare some of the ingredients. The kinks then describe
the colliding hadrons or ions, while the quasiparticles describe the formed
``plasma''. After the collision, most of the energy remains in the receding
kink-antikink, while the resulting field shows a central plateau, flat in
space-time
rapidity, indicating hydrodynamic scaling. Its height only weakly depends on
the initial speed of the kinks and scales with the coupling in a very
straightforward way. 

We found that the central region first has a finite speed of sound, of about
$0.8$ for $\lambda/m^2=1/1.25$ and about $0.4$ for $\lambda/m^2=1/12$. After a
short while this goes over in a free expansion, with vanishing pressure and
speed of sound.

In this aspect the qualitative difference between the Hartree and classical approximations is small.
Although it is difficult to compare the two at strong coupling at the later
stage, as the Hartree approximation is ``haunted'' by the ``symmetric minimum''
at $\phi=0$, at the early stage a meaningful comparison can be made. It shows
very similar results between the Hartree and classical approximations. At
smaller coupling the two are very similar even during the later stage, the
Hartree result shows somewhat more interaction at times around $50-100$ and
related
to
that a somewhat smaller energy density.

It would be interesting to have some handle on the error
induced by
the first order phase transition predicted by the Hartree approximation.
Although the lack of a systematic expansion parameter
in this approximation
makes it difficult to make
a quantitative
estimate, the above results seem to indicate a
qualitatively small influence,
even at strong coupling,
as long as the field is not moving too close to
the artificial ``symmetric minimum''.
However, at the later stage
in the strong coupling simulation of Fig.~\ref{fig:plasmalinehartree},
the field is actually fluctuating around this artificial minimum at $\phi=0$ and
the results can no longer be trusted. Similarly, it is not expected that the
Hartree approximation can accurately describe the dynamics of going through the
phase transition, see Ref.~\cite{SaSm00b}.

In Ref.~\cite{BeCo01} the Hartree ``plasma'' was also found to behave as a
``relativistic plasma,'' with speed of sound close to $1$, similar to what we
found
for $tm\lesssim8$.
However the result in \cite{BeCo01} was
obtained from a disintegrating Gaussian wave packet in the ``symmetric phase,''
in which the effective coupling is much
smaller\cite{SaSm02}, instead of a colliding
kink-antikink in the ``broken phase.'' No free expanding phase was found in that
reference.

Finally, we have briefly looked at the connection between a thermal
Bose-Ein\-stein distribution and the creation and annihilation of kinks in the
system. We found that one should only consider the complete field in the
description of kinks, not just the mean field. This means that we have to look
at quantum and ensemble averages only; the separate realisations give some
impression of what is happening but do not describe the full theory. It is
encouraging to see that although the kinks disappear from the mean field, our
rough kink indicator does not go to zero, but becomes constant, i.e., the
creation and annihilation rates become equal. However, we do find fewer pairs in
the Hartree approximation than in the classical theory. This might be related to
the higher instability of Hartree kinks, which can radiate quantum particles, in
the Hartree description, more energy is carried by radiation than in the
classical theory.

\acknowledgments{
I would like to thank Jan Smit, without whom this article would not exist. M.S.
was financially supported by the Stichting FOM.
}

\bibliography{liter}

\begin{thebibliography}{37}
\expandafter\ifx\csname natexlab\endcsname\relax\def\natexlab#1{#1}\fi
\expandafter\ifx\csname bibnamefont\endcsname\relax
  \def\bibnamefont#1{#1}\fi
\expandafter\ifx\csname bibfnamefont\endcsname\relax
  \def\bibfnamefont#1{#1}\fi
\expandafter\ifx\csname citenamefont\endcsname\relax
  \def\citenamefont#1{#1}\fi
\expandafter\ifx\csname url\endcsname\relax
  \def\url#1{\texttt{#1}}\fi
\expandafter\ifx\csname urlprefix\endcsname\relax\def\urlprefix{URL }\fi
\providecommand{\bibinfo}[2]{#2}
\providecommand{\eprint}[2][]{\url{#2}}

\bibitem[{\citenamefont{Aarts et~al.}(2002)\citenamefont{Aarts, Ahrensmeier,
  Baier, Berges, and Serreau}}]{AaAh02}
\bibinfo{author}{\bibfnamefont{G.}~\bibnamefont{Aarts}},
  \bibinfo{author}{\bibfnamefont{D.}~\bibnamefont{Ahrensmeier}},
  \bibinfo{author}{\bibfnamefont{R.}~\bibnamefont{Baier}},
  \bibinfo{author}{\bibfnamefont{J.}~\bibnamefont{Berges}}, \bibnamefont{and}
  \bibinfo{author}{\bibfnamefont{J.}~\bibnamefont{Serreau}},
  \bibinfo{journal}{Phys. Rev.} \textbf{\bibinfo{volume}{D66}},
  \bibinfo{pages}{045008} (\bibinfo{year}{2002}), \eprint{hep-ph/0201308}.

\bibitem[{\citenamefont{Cooper et~al.}(2003)\citenamefont{Cooper, Dawson, and
  Mihaila}}]{CoDa02}
\bibinfo{author}{\bibfnamefont{F.}~\bibnamefont{Cooper}},
  \bibinfo{author}{\bibfnamefont{J.~F.} \bibnamefont{Dawson}},
  \bibnamefont{and} \bibinfo{author}{\bibfnamefont{B.}~\bibnamefont{Mihaila}},
  \bibinfo{journal}{Phys. Rev.} \textbf{\bibinfo{volume}{D67}},
  \bibinfo{pages}{056003} (\bibinfo{year}{2003}), \eprint{hep-ph/0209051}.

\bibitem[{\citenamefont{Sall{\'e} et~al.}(2001)\citenamefont{Sall{\'e}, Smit,
  and Vink}}]{SaSm00a}
\bibinfo{author}{\bibfnamefont{M.}~\bibnamefont{Sall{\'e}}},
  \bibinfo{author}{\bibfnamefont{J.}~\bibnamefont{Smit}}, \bibnamefont{and}
  \bibinfo{author}{\bibfnamefont{J.~C.} \bibnamefont{Vink}},
  \bibinfo{journal}{Phys. Rev.} \textbf{\bibinfo{volume}{D64}},
  \bibinfo{pages}{025016} (\bibinfo{year}{2001}),
  \eprint[http://arXiv.org/abs]{hep-ph/0012346}.

\bibitem[{\citenamefont{Sall{\'e} et~al.}(2002)\citenamefont{Sall{\'e}, Smit,
  and Vink}}]{SaSm00b}
\bibinfo{author}{\bibfnamefont{M.}~\bibnamefont{Sall{\'e}}},
  \bibinfo{author}{\bibfnamefont{J.}~\bibnamefont{Smit}}, \bibnamefont{and}
  \bibinfo{author}{\bibfnamefont{J.~C.} \bibnamefont{Vink}},
  \bibinfo{journal}{Nucl. Phys.} \textbf{\bibinfo{volume}{B625}},
  \bibinfo{pages}{495} (\bibinfo{year}{2002}),
  \eprint[http://arXiv.org/abs]{hep-ph/0012362}.

\bibitem[{\citenamefont{Sall{\'e} and Smit}(2003)}]{SaSm02}
\bibinfo{author}{\bibfnamefont{M.}~\bibnamefont{Sall{\'e}}} \bibnamefont{and}
  \bibinfo{author}{\bibfnamefont{J.}~\bibnamefont{Smit}},
  \bibinfo{journal}{Phys. Rev.} \textbf{\bibinfo{volume}{D67}},
  \bibinfo{pages}{116006} (\bibinfo{year}{2003}),
  \eprint[http://arXiv.org/abs]{hep-ph/0208139}.

\bibitem[{\citenamefont{Rajaraman}(1982)}]{Ra82}
\bibinfo{author}{\bibfnamefont{R.}~\bibnamefont{Rajaraman}},
  \emph{\bibinfo{title}{Solitons and Instantons}}
  (\bibinfo{publisher}{North-Holland Publishing Company},
  \bibinfo{address}{Amsterdam}, \bibinfo{year}{1982}).

\bibitem[{\citenamefont{Bogomol'nyi}(1976)}]{Bo76}
\bibinfo{author}{\bibfnamefont{E.~B.} \bibnamefont{Bogomol'nyi}},
  \bibinfo{journal}{Sov. J. Nucl. Phys.} \textbf{\bibinfo{volume}{24}},
  \bibinfo{pages}{449} (\bibinfo{year}{1976}).

\bibitem[{\citenamefont{Gleiser and Sornborger}(2000)}]{GlSo00}
\bibinfo{author}{\bibfnamefont{M.}~\bibnamefont{Gleiser}} \bibnamefont{and}
  \bibinfo{author}{\bibfnamefont{A.}~\bibnamefont{Sornborger}},
  \bibinfo{journal}{Phys. Rev.} \textbf{\bibinfo{volume}{E62}},
  \bibinfo{pages}{1368} (\bibinfo{year}{2000}),
  \eprint[http://arXiv.org/abs]{patt-sol/9909002}.

\bibitem[{\citenamefont{Speight and Ward}(1994)}]{SpWa94}
\bibinfo{author}{\bibfnamefont{J.~M.} \bibnamefont{Speight}} \bibnamefont{and}
  \bibinfo{author}{\bibfnamefont{R.}~\bibnamefont{Ward}},
  \bibinfo{journal}{Nonlinearity} \textbf{\bibinfo{volume}{7}},
  \bibinfo{pages}{475} (\bibinfo{year}{1994}).

\bibitem[{\citenamefont{Speight}(1997)}]{Sp97}
\bibinfo{author}{\bibfnamefont{J.~M.} \bibnamefont{Speight}},
  \bibinfo{journal}{Nonlinearity} \textbf{\bibinfo{volume}{10}},
  \bibinfo{pages}{1615} (\bibinfo{year}{1997}),
  \eprint[http://arXiv.org/abs]{patt-sol/9703005}.

\bibitem[{\citenamefont{Speight}(1999)}]{Sp99}
\bibinfo{author}{\bibfnamefont{J.~M.} \bibnamefont{Speight}},
  \bibinfo{journal}{Nonlinearity} \textbf{\bibinfo{volume}{12}},
  \bibinfo{pages}{1373} (\bibinfo{year}{1999}),
  \eprint[http://arXiv.org/abs]{hep-th/9812064}.

\bibitem[{\citenamefont{Adib and Almeida}(2001)}]{AdAl01}
\bibinfo{author}{\bibfnamefont{A.~B.} \bibnamefont{Adib}} \bibnamefont{and}
  \bibinfo{author}{\bibfnamefont{C.~A.~S.} \bibnamefont{Almeida}},
  \bibinfo{journal}{Phys. Rev.} \textbf{\bibinfo{volume}{E64}},
  \bibinfo{pages}{37701} (\bibinfo{year}{2001}),
  \eprint[http://arXiv.org/abs]{hep-th/0104225}.

\bibitem[{\citenamefont{Boyanovsky et~al.}(1998)\citenamefont{Boyanovsky,
  Cooper, de~Vega, and Sodano}}]{BoCo98}
\bibinfo{author}{\bibfnamefont{D.}~\bibnamefont{Boyanovsky}},
  \bibinfo{author}{\bibfnamefont{F.}~\bibnamefont{Cooper}},
  \bibinfo{author}{\bibfnamefont{H.~J.} \bibnamefont{de~Vega}},
  \bibnamefont{and} \bibinfo{author}{\bibfnamefont{P.}~\bibnamefont{Sodano}},
  \bibinfo{journal}{Phys. Rev.} \textbf{\bibinfo{volume}{D58}},
  \bibinfo{pages}{025007} (\bibinfo{year}{1998}), \eprint{hep-ph/9802277}.

\bibitem[{\citenamefont{Dashen et~al.}(1974)\citenamefont{Dashen, Hasslacher,
  and Neveu}}]{DaHa74}
\bibinfo{author}{\bibfnamefont{R.~F.} \bibnamefont{Dashen}},
  \bibinfo{author}{\bibfnamefont{B.}~\bibnamefont{Hasslacher}},
  \bibnamefont{and} \bibinfo{author}{\bibfnamefont{A.}~\bibnamefont{Neveu}},
  \bibinfo{journal}{Phys. Rev.} \textbf{\bibinfo{volume}{D10}},
  \bibinfo{pages}{4130} (\bibinfo{year}{1974}).

\bibitem[{\citenamefont{Alonso~Izquierdo
  et~al.}(2002)\citenamefont{Alonso~Izquierdo, Garc{\'\i}a~Fuertes,
  Gonz{\'a}lez~Le{\'o}n, and Mateos~Guilarte}}]{AlGa02}
\bibinfo{author}{\bibfnamefont{A.}~\bibnamefont{Alonso~Izquierdo}},
  \bibinfo{author}{\bibfnamefont{W.}~\bibnamefont{Garc{\'\i}a~Fuertes}},
  \bibinfo{author}{\bibfnamefont{M.~A.} \bibnamefont{Gonz{\'a}lez~Le{\'o}n}},
  \bibnamefont{and}
  \bibinfo{author}{\bibfnamefont{J.}~\bibnamefont{Mateos~Guilarte}},
  \bibinfo{journal}{Nucl. Phys.} \textbf{\bibinfo{volume}{B635}},
  \bibinfo{pages}{525} (\bibinfo{year}{2002}),
  \eprint[http://arXiv.org/abs]{hep-th/0201084}.

\bibitem[{\citenamefont{Weidig}(1999)}]{We99}
\bibinfo{author}{\bibfnamefont{T.}~\bibnamefont{Weidig}}
  (\bibinfo{year}{1999}), \eprint{hep-th/9912005}.

\bibitem[{\citenamefont{de~Vega}(1976)}]{Ve76}
\bibinfo{author}{\bibfnamefont{H.~J.} \bibnamefont{de~Vega}},
  \bibinfo{journal}{Nucl. Phys.} \textbf{\bibinfo{volume}{B115}},
  \bibinfo{pages}{411} (\bibinfo{year}{1976}).

\bibitem[{\citenamefont{Verwaest}(1977)}]{Ve77}
\bibinfo{author}{\bibfnamefont{J.}~\bibnamefont{Verwaest}},
  \bibinfo{journal}{Nucl. Phys.} \textbf{\bibinfo{volume}{B123}},
  \bibinfo{pages}{100} (\bibinfo{year}{1977}).

\bibitem[{\citenamefont{Ciria and Taranc{\'o}n}(1994)}]{CiTa94}
\bibinfo{author}{\bibfnamefont{J.~C.} \bibnamefont{Ciria}} \bibnamefont{and}
  \bibinfo{author}{\bibfnamefont{A.}~\bibnamefont{Taranc{\'o}n}},
  \bibinfo{journal}{Phys. Rev.} \textbf{\bibinfo{volume}{D49}},
  \bibinfo{pages}{1020} (\bibinfo{year}{1994}),
  \eprint[http://arXiv.org/abs]{hep-lat/9309019}.

\bibitem[{\citenamefont{Ardekani and Williams}(1999)}]{ArWi99}
\bibinfo{author}{\bibfnamefont{A.}~\bibnamefont{Ardekani}} \bibnamefont{and}
  \bibinfo{author}{\bibfnamefont{A.~G.} \bibnamefont{Williams}},
  \bibinfo{journal}{Austral. J. Phys.} \textbf{\bibinfo{volume}{52}},
  \bibinfo{pages}{929} (\bibinfo{year}{1999}),
  \eprint[http://arXiv.org/abs]{hep-lat/9811002}.

\bibitem[{\citenamefont{Kadanoff}(1969)}]{Ka69}
\bibinfo{author}{\bibfnamefont{L.~P.} \bibnamefont{Kadanoff}},
  \bibinfo{journal}{Phys. Rev. Lett.} \textbf{\bibinfo{volume}{23}},
  \bibinfo{pages}{1430} (\bibinfo{year}{1969}).

\bibitem[{\citenamefont{Kadanoff and Ceva}(1971)}]{KaCe71}
\bibinfo{author}{\bibfnamefont{L.~P.} \bibnamefont{Kadanoff}} \bibnamefont{and}
  \bibinfo{author}{\bibfnamefont{H.}~\bibnamefont{Ceva}},
  \bibinfo{journal}{Phys. Rev.} \textbf{\bibinfo{volume}{B3}},
  \bibinfo{pages}{3918} (\bibinfo{year}{1971}).

\bibitem[{\citenamefont{Bergner and Bettencourt}(2003)}]{BeBe03}
\bibinfo{author}{\bibfnamefont{Y.}~\bibnamefont{Bergner}} \bibnamefont{and}
  \bibinfo{author}{\bibfnamefont{L.~M.~A.} \bibnamefont{Bettencourt}}
  (\bibinfo{year}{2003}), \eprint{hep-th/0305190}.

\bibitem[{\citenamefont{Campbell et~al.}(1983)\citenamefont{Campbell,
  Schonfeld, and Wingate}}]{CaSc83}
\bibinfo{author}{\bibfnamefont{D.~K.} \bibnamefont{Campbell}},
  \bibinfo{author}{\bibfnamefont{J.~F.} \bibnamefont{Schonfeld}},
  \bibnamefont{and} \bibinfo{author}{\bibfnamefont{C.~A.}
  \bibnamefont{Wingate}}, \bibinfo{journal}{Physica}
  \textbf{\bibinfo{volume}{D9}}, \bibinfo{pages}{1} (\bibinfo{year}{1983}).

\bibitem[{\citenamefont{Anninos et~al.}(1991)\citenamefont{Anninos, Oliveira,
  and Matzner}}]{AnOl91}
\bibinfo{author}{\bibfnamefont{P.}~\bibnamefont{Anninos}},
  \bibinfo{author}{\bibfnamefont{S.}~\bibnamefont{Oliveira}}, \bibnamefont{and}
  \bibinfo{author}{\bibfnamefont{R.~A.} \bibnamefont{Matzner}},
  \bibinfo{journal}{Phys. Rev.} \textbf{\bibinfo{volume}{D44}},
  \bibinfo{pages}{1147} (\bibinfo{year}{1991}).

\bibitem[{\citenamefont{Segur and Kruskal}(1987)}]{SeKr87}
\bibinfo{author}{\bibfnamefont{H.}~\bibnamefont{Segur}} \bibnamefont{and}
  \bibinfo{author}{\bibfnamefont{M.~D.} \bibnamefont{Kruskal}},
  \bibinfo{journal}{Phys. Rev. Lett.} \textbf{\bibinfo{volume}{58}},
  \bibinfo{pages}{747} (\bibinfo{year}{1987}).

\bibitem[{\citenamefont{Geicke}(1994)}]{Ge94}
\bibinfo{author}{\bibfnamefont{J.}~\bibnamefont{Geicke}},
  \bibinfo{journal}{Phys. Rev.} \textbf{\bibinfo{volume}{E49}},
  \bibinfo{pages}{3539} (\bibinfo{year}{1994}), \bibinfo{note}{brief Reports}.

\bibitem[{\citenamefont{Bjorken}(1983)}]{Bj83}
\bibinfo{author}{\bibfnamefont{J.}~\bibnamefont{Bjorken}},
  \bibinfo{journal}{Phys. Rev.} \textbf{\bibinfo{volume}{D27}},
  \bibinfo{pages}{140} (\bibinfo{year}{1983}).

\bibitem[{\citenamefont{Back et~al.}(2002)}]{Ba01}
\bibinfo{author}{\bibfnamefont{B.~B.} \bibnamefont{Back}} \bibnamefont{et~al.}
  (\bibinfo{collaboration}{PHOBOS}), \bibinfo{journal}{Phys. Rev. Lett.}
  \textbf{\bibinfo{volume}{88}}, \bibinfo{pages}{022302}
  (\bibinfo{year}{2002}), \eprint{nucl-ex/0108009}.

\bibitem[{\citenamefont{Bettencourt et~al.}(2002)\citenamefont{Bettencourt,
  Cooper, and Pao}}]{BeCo01}
\bibinfo{author}{\bibfnamefont{L.~M.~A.} \bibnamefont{Bettencourt}},
  \bibinfo{author}{\bibfnamefont{F.}~\bibnamefont{Cooper}}, \bibnamefont{and}
  \bibinfo{author}{\bibfnamefont{K.}~\bibnamefont{Pao}},
  \bibinfo{journal}{Phys. Rev. Lett.} \textbf{\bibinfo{volume}{89}},
  \bibinfo{pages}{112301} (\bibinfo{year}{2002}), \eprint{hep-ph/0109108}.

\bibitem[{\citenamefont{Currie et~al.}(1980)\citenamefont{Currie, Krumhansl,
  Bishop, and Trullinger}}]{CuKr80}
\bibinfo{author}{\bibfnamefont{J.~F.} \bibnamefont{Currie}},
  \bibinfo{author}{\bibfnamefont{J.~A.} \bibnamefont{Krumhansl}},
  \bibinfo{author}{\bibfnamefont{A.~R.} \bibnamefont{Bishop}},
  \bibnamefont{and} \bibinfo{author}{\bibfnamefont{S.~E.}
  \bibnamefont{Trullinger}}, \bibinfo{journal}{Phys. Rev.}
  \textbf{\bibinfo{volume}{B22}}, \bibinfo{pages}{477} (\bibinfo{year}{1980}).

\bibitem[{\citenamefont{Grigoriev and Rubakov}(1988)}]{GrRu88}
\bibinfo{author}{\bibfnamefont{D.~Y.} \bibnamefont{Grigoriev}}
  \bibnamefont{and} \bibinfo{author}{\bibfnamefont{V.~A.}
  \bibnamefont{Rubakov}}, \bibinfo{journal}{Nucl. Phys.}
  \textbf{\bibinfo{volume}{B299}}, \bibinfo{pages}{67} (\bibinfo{year}{1988}).

\bibitem[{\citenamefont{Alford et~al.}(1992)\citenamefont{Alford, Feldman, and
  Gleiser}}]{AlFe92}
\bibinfo{author}{\bibfnamefont{M.~G.} \bibnamefont{Alford}},
  \bibinfo{author}{\bibfnamefont{H.}~\bibnamefont{Feldman}}, \bibnamefont{and}
  \bibinfo{author}{\bibfnamefont{M.}~\bibnamefont{Gleiser}},
  \bibinfo{journal}{Phys. Rev. Lett.} \textbf{\bibinfo{volume}{68}},
  \bibinfo{pages}{1645} (\bibinfo{year}{1992}).

\bibitem[{\citenamefont{Alexander and Habib}(1993)}]{AlHa93a}
\bibinfo{author}{\bibfnamefont{F.~J.} \bibnamefont{Alexander}}
  \bibnamefont{and} \bibinfo{author}{\bibfnamefont{S.}~\bibnamefont{Habib}},
  \bibinfo{journal}{Phys. Rev. Lett.} \textbf{\bibinfo{volume}{71}},
  \bibinfo{pages}{955} (\bibinfo{year}{1993}),
  \eprint[http://arXiv.org/abs]{hep-th/9212059}.

\bibitem[{\citenamefont{Alexander et~al.}(1993)\citenamefont{Alexander, Habib,
  and Kovner}}]{AlHa93b}
\bibinfo{author}{\bibfnamefont{F.~J.} \bibnamefont{Alexander}},
  \bibinfo{author}{\bibfnamefont{S.}~\bibnamefont{Habib}}, \bibnamefont{and}
  \bibinfo{author}{\bibfnamefont{A.}~\bibnamefont{Kovner}},
  \bibinfo{journal}{Phys. Rev.} \textbf{\bibinfo{volume}{E48}},
  \bibinfo{pages}{4284} (\bibinfo{year}{1993}),
  \eprint[http://arXiv.org/abs]{hep-th/9308103}.

\bibitem[{\citenamefont{Habib and Lythe}(2000)}]{HaLy00}
\bibinfo{author}{\bibfnamefont{S.}~\bibnamefont{Habib}} \bibnamefont{and}
  \bibinfo{author}{\bibfnamefont{G.}~\bibnamefont{Lythe}},
  \bibinfo{journal}{Phys. Rev. Lett.} \textbf{\bibinfo{volume}{84}},
  \bibinfo{pages}{1070} (\bibinfo{year}{2000}),
  \eprint[http://arXiv.org/abs]{cond-mat/9911228}.

\bibitem[{\citenamefont{Holzwarth}(2003)}]{Ho03}
\bibinfo{author}{\bibfnamefont{G.}~\bibnamefont{Holzwarth}},
  \bibinfo{journal}{Phys. Rev.} \textbf{\bibinfo{volume}{D68}},
  \bibinfo{pages}{016008} (\bibinfo{year}{2003}), \eprint{hep-ph/0303208}.

\end{thebibliography}

\end{document}